\documentclass{llncs}

\usepackage{amsmath}
\usepackage{amssymb} 
\usepackage{multirow}
\usepackage{epic,eepic,latexsym}

\usepackage{gastex} 
\usepackage{url}

\newcommand{\set}[1]{\{ #1 \}}

\newcommand{\pair}[2]{\langle #1,#2 \rangle}
\newcommand{\triple}[3]{\langle #1,#2,#3 \rangle}

\newcommand{\Nat}{\mathbb{N}}

\newcommand{\powerset}[1]{{\cal P}(#1)}

\newcommand{\length}[1]{|#1|}
\newcommand{\GF}{{\rm GF}} 
\newcommand{\FO}{{\rm FO}}

\newcommand{\aformula}{\phi}
\newcommand{\aformulabis}{\psi}
\newcommand{\aformulater}{\varphi}
\newcommand{\aconst}{{\rm c}}
\newcommand{\bo}[1]{[#1]}
\newcommand{\diam}[1]{\langle #1 \rangle}

\newcommand{\subf}[1]{sub(#1)}

\newcommand{\varprop}{{\rm PROP}}

\newcommand{\avarprop}{{\rm p}}

\newcommand{\anominal}{\mathbf i}

\newcommand{\amodel}{{\cal M}}
\newcommand{\aframe}{{\cal F}}
\newcommand{\alanguage}{{\rm L}}
\newcommand{\alang}{{\rm L}}
\newcommand{\asystem}{{\rm S}}

\newcommand{\astring}{u}
\newcommand{\astringbis}{v}

\newcommand{\aalphabet}{\Sigma}

\newcommand{\aletter}{a}
\newcommand{\aletterbis}{b}
\newcommand{\aletterter}{c}

\newcommand{\statepred}[1]{{\mathbf P}_{#1}}
\newcommand{\relpred}[1]{{\mathbf R}_{#1}}
\newcommand{\proppred}[1]{{\mathbf p}}
\newcommand{\nompred}[1]{{\mathbf c}_{#1}}

\newcommand{\astate}{q}

\newcommand{\aautomaton}{\mathcal{A}}

\newcommand{\egdef}{\stackrel{\mbox{\begin{tiny}def\end{tiny}}}{=}} 

\newcommand{\exptime}{{\sc EXPTIME}}
\newcommand{\nexptime}{{\sc NEXPTIME}}
\newcommand{\twoexptime}{{\sc 2EXPTIME}}

\newcommand{\pspace}{{\sc PSPACE}}

\begin{document}

\pagestyle{plain}
\pagenumbering{arabic}


\title{
       Deciding regular grammar logics with converse through first-order 
       logic\thanks{Revised version of~\cite{Demri&deNivelle03}.}
      }

\author{St\'ephane Demri\inst{1}  and Hans de Nivelle\inst{2}}

\institute{LSV/CNRS UMR 8643 \& INRIA Futurs projet SECSI \& ENS Cachan \\
           61, av. Pdt. Wilson, 94235 Cachan Cedex, France                         \\
           email: {\tt demri@lsv.ens-cachan.fr}
           \and
           Max Planck Institut f\"ur Informatik \\
           Stuhlsatzenhausweg 85 \\
           66123 Saarbr\"ucken, Germany \\
           email: {\tt nivelle@mpi-sb.mpg.de}
          }

\maketitle

\date{November 18th, 2003}


\begin{abstract} 
   We provide a simple translation of the satisfiability problem
   for regular grammar logics with converse into $\GF^2$, which is the 
   intersection of 
   the guarded fragment and the $ 2 $-variable fragment 
   of first-order logic. 
   This translation is theoretically interesting because it translates
   modal logics with certain frame conditions into first-order logic,
   without explicitly expressing the frame conditions.
   
   A consequence of the translation is that the general satisfiability 
   problem for regular grammar logics with converse is in \exptime.
   This extends a previous result of the first author 
   for grammar logics without converse.
   Using the same method, we show how some other modal logics can 
   be naturally translated into
   $ \GF^2, $ including nominal tense logics and intuitionistic
   logic.  
   In our view, the results in this paper show that the natural 
   first-order fragment corresponding to regular grammar logics is simply 
   $ \GF^2 $ without extra machinery such as fixed point-operators.
\end{abstract}


\section{Introduction}
\label{section-introduction}

\paragraph{Translating modal logics.}
Modal logics are used in many areas of Computer Science, as for example
knowledge representation, model-checking, and temporal reasoning.
For {\em theorem proving} in modal logics, two main
approaches can be distinguished. The first approach is
to develope a theorem
prover directly for the logic under consideration.
The second approach is to translate the logic into some general
logic, usually first-order logic.
The first approach has the advantage that a specialized
algorithm can make use of specific properties of the logic under
consideration, enabling optimizations that do not work in
general. In addition, implementation of a single modal logic
is usually easier than implementing full first-order logic.
But on the other hand, there are many modal logics,
and it is simply not feasible to construct optimized
theorem provers for all of them.
The advantage of the second approach is that only one theorem
prover needs to be written which can be reused for
all translatable modal logics.
In addition, a translation method can be expected to be more robust
against small changes in the logic.
Therefore translating seems to be the more sensible approach for most
modal logics, with the exception of a few main ones.

Translation of modal logics into first-order logic,
with the explicit goal to mechanise such logics
is an approach that has been introduced in \cite{Morgan76}.
Morgan distinguishes two types of translations:
The {\em semantical translation}, which is nowadays known as the
{\em relational translation} (see e.g.,
\cite{Fine75b,vanBenthem76,Moore77}) and the {\em syntactic} translation,
which consists in reifing modal formulae (i.e., transforming them
into first-order terms) and in translating the axioms
and inference rules from a Hilbert-style system into classical logic using
an additional provability predicate symbol.
This is also sometimes called {\em reflection}.
With such a syntactic translation, every propositional normal modal
logic with a finite axiomatization 
can be translated into classical predicate logic.
However, using this general translation, decidability of
modal logics is lost.
Therefore we will study relational translations in this paper,
which instead of simply translating a modal formula into
full first-order logic, can translate modal formulas into a
decidable subset of first-order logic.
The fragment that we will be using is $ \GF^2, $ the intersection of
the $ 2 $-variable fragment and the guarded fragment
We modify the relational translation in such a way that explicit
translation of frame properties can be avoided. In this way, 
many modal logics with frame properties outside $ \GF^2 $
can be translated into $ \GF^2. $ 

A survey on translation methods for modal logics can be found in
\cite{Ohlbachetal01}, where more references are provided, for instance about
the functional translation (see e.g.,
\cite{Herzig89,Ohlbach93b,Nonnengart96}),
see also in~\cite{Orlowska88b,DAgostino&Montanari&Policriti95b}
for other types of translations.

\paragraph{Guarded fragments.}
Both the guarded fragment, introduced in \cite{Andreka&Nemeti&vanBenthem98}
(see also~\cite{deNivelle98,Ganzinger&deNivelle99,deNivelle&Schmidt&Hustadt00,deNivelle&deRijke03}) and 
$ \FO^2, $ the fragment of classical logic with two variables 
\cite{gabb:expr81,Gradel&Kolaitis&Vardi97,deNivelle&PrattHartmann01},
have been used for the purpose of 'hosting' translations of
modal formulas. 
The authors of \cite{Andreka&Nemeti&vanBenthem98} explicitly mention
the goal of identifying 'the modal fragment of first-order logic' as
a motivation for introducing the guarded fragment. 
Apart from having nice logical properties 
\cite{Andreka&Nemeti&vanBenthem98}, the guarded fragment $ \GF $ has 
an \exptime-complete satisfiability problem when the maximal arity 
of the predicate symbols is fixed in advance~\cite{Gradel99c}. 
Hence its worst-case 
complexity is identical to some simple extensions of modal logic K, 
as for example K augmented with universal modality \cite{Spaan93}.
Moreover, mechanisation of the guarded fragment is possible thanks
to the design of efficient resolution-based decision procedures 
\cite{deNivelle98,Ganzinger&deNivelle99}.
In \cite{GF_Hladik_tableaux}, a tableau procedure for the guarded fragment
with equality based on~\cite{Hirsch&Tobies00} is implemented and tested;
see also a prover for $ \FO^2 $ described in~\cite{Marx&Mikulas&Schlobach99}.

However, there are some simple modal logics with the 
satisfiability problem in \pspace \ (\cite{Ladner77}) that 
cannot be translated into $ \GF $ through the relational translation. 
The reason for
this is the fact that the frame condition that characterizes the logic 
cannot be expressed in $ \GF. $ 
The simplest example of such a logic is probably S4 which
is characterized by transitivity (many other examples
will be given throughout the paper). 
Adding transitivity axioms to a 
$ \GF $-formula causes undecidability. (See \cite{Gradel99b})

Because of the apparent insufficiency of $ \GF $ to capture 
basic modal logics, various extensions of $ \GF $ have 
been proposed and studied. In \cite{Ganzinger&Meyer&Veanes99},
it was shown that $ \GF^2 $ with transitivity axioms is decidable, 
on the condition that binary predicates occur only in guards. 
The complexity bound given there is non-elementary, which
makes the fragment not useful for practical purposes.

In \cite{Szwast_Tendera01}, the complexity bound for $ \GF^2 $ with 
transitive guards is improved to \twoexptime \ 
and it was shown \twoexptime-hard in \cite{Kieronski03}. 
As a consequence, the resulting strategy is not the
most efficient strategy to mechanize modal logics
with transitive relations (such as S4) 

Another fragment was explored in 
\cite{Gradel&Walukiewicz99}, see also \cite{Gradel99b}.
There it was shown that $ \mu \GF, $ the guarded fragment 
extended with a $\mu$-calculus-style fixed point operator is still
decidable and in \twoexptime. 
This fragment does contain the simple modal logic S4, but the machinery
is much more heavy than the 
than a direct decision procedure would be.  
After all, there exist simple tableaux procedures for S4.
In addition, $ \mu \GF $ does not have the finite model property,
although S4 has.

\paragraph{Almost structure-preserving translations.}

In this paper, we put emphasis on the fact that $ \GF^2 $ is a sufficiently
well-designed fragment of classical logic for dealing with a 
large variety of modal logics.
An approach that seems better suited for theorem proving, and
that does more justice to the low complexities of simple
modal logics is the approach taken 
in \cite{deNivelle99,deNivelle01}. There an
almost structure-preserving translation
from the modal logics S4, S5 and K5 into $ \GF^2 $ was given.
The subformulas of a modal formula are translated in the standard way, 
except for subformulas which are formed by a universal modality.
Universal subformulas are translated into a sequence of formulas, 
the exact form of which is determined by the frame condition. 
In \cite{deNivelle99,deNivelle01}, the translations and their correctness 
proofs were ad hoc, and it was not
clear upon which principles they are based.
In this paper we show that the almost structure preserving translation 
relies on the fact that
the frame conditions for K4, S4 and K5 are {\em regular} in some
sense that will be made precise in Section~\ref{section-preliminaries}.
The simplicity of the almost structure-preserving translation
leaves hope that $ \GF^2 $ may be 
rich enough after all to naturally capture most of the basic modal logics.

We call the translation method almost structure-preserving
because it preserves the structure of the formula almost
completely. Only for subformulas of form $ \bo{\aletter} \aformula $ 
does the translation differ from the usual relational translation. 
On these subformulas, a sequence of formulas is generated
that simulates an NDFA based on the frame condition of the modal logic. 
In our view this translation also provides an explanation 
why some modal logics like S4, have nice tableau procedures (see 
e.g.~\cite{Heuerding&Seyfried&Zimmermann96,Gore99,Massacci00,Farinas&Gasquet02,Horrocks&Sattler03}):
The tableau rule for subformulas of form $ \bo{\aletter} \aformula $ 
can be viewed as simulating an NDFA, in the same way as the 
almost structure-preserving translation. 

\paragraph{Our contribution.}
We show that the methods of \cite{deNivelle99} can
be extended to a very large class of modal logics. Some of the modal
logics in this class have frame properties that can be expressed
only by recursive conditions, like for example transitivity.
By a {\em recursive condition} we mean a condition that needs to be
iterated in order to reach a fixed point. 
The class of modal logics that we consider is the class of 
{\em regular grammar logics with converse}.
The axioms of such modal logics are of form 
$\bo{\aletter_0} \avarprop \Rightarrow \bo{\aletter_1} \ldots \bo{\aletter_n} 
\avarprop$ where each $ \bo{a_i} $ is either a forward or
a backward modality. Another condition called {\em regularity} is required
and will be formally defined in Section~\ref{section-preliminaries}.

With our translation, we are able to translate numerous
modal logics into $ \GF^2 = \FO^2 \cap \GF, $ despite
the fact that their frame conditions are not expressible
in $ \FO^2 \cup \GF $.
These logics include the standard modal logics K4, S4, K5, K45, S5,
some information logics (see e.g.~\cite{Vakarelov87}),
nominal tense logics (see e.g.~\cite{Areces&Blackburn&Marx00}),
description logics (see e.g.~\cite{Sattler96,Horrocks&Sattler99}),
propositional intuitionistic logic (see e.g.~\cite{Chagrov&Zakharyaschev97})
and bimodal logics for intuitionistic modal logics
{\bf IntK}$_{\Box} + \Gamma$
 as those considered 
in~\cite{Wolter&Zakharyaschev97}.
Hence the main contribution of the paper is the design of a very simple and generic translation from
regular grammar logics with converse into the decidable fragment of classical logic $\GF^2$.
The translation is easy to implement and 
it mimics the behavior of some tableaux-based calculi for modal logics.
As a consequence, we are able to show that the source logics that can be translated into
$ \GF^2 $ have a satisfiability problem in \exptime. This allows 
us to establish such an upper bound uniformly
for a very large class of modal logics, for instance for intuitionistic modal logics 
(another approach is followed in~\cite{Alechina&Shkatov03} leading to a bit
less tight complexity upper bound).
We are considering here the satisfiability problem. However because of the very nature
of the regular grammar logics with converse, our results apply also to the global satisfiability problem
and to the logical consequence problem. By contrast, this work does not deal with the  model-checking problem since
this problem amounts to a subproblem of the model-checking problem for classical FOL known to be
\pspace-complete.

We do not claim that for most source logics the existence of a transformation into $\GF^2$ of low complexity is 
very surprising. 
In fact it is easy to see that from each simple modal logic for instance in 
\pspace \ there must exist a polynomial transformation into $\GF^2$, because
\pspace \ is a subclass of \exptime.
The \exptime-completeness of fixed-arity $\GF$ implies that there
exists a polynomial time transformation from every logic in 
\pspace \ into fixed-arity $\GF$. 
(It can even be shown that there exists a logarithmic space transformation.)
However, the translation that establishes the reduction
would normally make use of first principles on 
Turing machines. Trying to efficiently decide modal logics through such a 
transformation would amount to finding an optimal implementation
in Turing machines, which is no easier than  a direct 
implementation on a standard computer. 
Although many translated regular grammar logics with converse are
\exptime-hard, see e.g.~\cite{Demri01}, our translation is optimal
for the logics in \pspace, and it is part of our future work to 
look for fragments of $\GF$ which are in \pspace.

%

Our paper answers a question stated in~\cite{Demri01}: 
Is there a decidable first-order fragment, into which the regular
grammar logics can be translated in a natural way?
The translation method that we give in this paper suggests
that $ \GF^2 $ is the answer. 
It is too early to state that the transformation from 
regular grammar logics with converse into $\GF^2$
defined in the paper can be used to mechanize efficiently such source logics
with a prover for $\GF^2$, but we have shown evidence that $\GF^2$ is a most 
valuable decidable first-order fragment to translate modal logics into, even 
though their frame conditions are not expressible in $\GF^2$

\paragraph{Related work.}
Complexity issues for regular grammar logics have been
studied in \cite{Demri01,Demri01b}
(see also \cite{Baldoni98,Baldoni&Giordano&Martelli98}) whereas grammar logics
are introduced in~\cite{Farinas&Penttonen88}. 
Frame conditions involving the converse
relations are not treated in \cite{Demri01,Demri01b}.
These are needed for example for S5 modal connectives. 
The current work can be viewed as a natural
continuation
of \cite{deNivelle99} and \cite{Demri01}. 
In this paper, we use a direct translation into decidable
first-order fragments instead of a translation into propositional dynamic
logic as done in \cite{Demri01} (see details about the latter approach in 
Section~\ref{section-alternative}). \\
The frame conditions considered in the present work can be defined by the
MSO definable closure operators~\cite{Ganzinger&Meyer&Veanes99}.
However, it is worth noting that by contrast to what is done in
\cite{Ganzinger&Meyer&Veanes99}, we obtain the optimal complexity 
upper bound for
the class of regular grammar logics with converse (\exptime) since
the first-order fragment we consider is much more restricted than the 
one in~\cite{Ganzinger&Meyer&Veanes99}. Moreover, we do not use MSO 
definable built-in
relations, just plain GF$^2$.

The recent work~\cite{Hustadt&Schmidt03} presents another translation of modal
logics into decidable fragments of classical logic by encoding adequately the
modal
theories (see also~\cite{Ohlbach98}).
Although for logics such as S4, the method in~\cite{Hustadt&Schmidt03}
is very similar to ours, it is still open how the methods are related
in the general case. For instance, no regularity conditions are explicitly
involved in~\cite{Hustadt&Schmidt03}
whereas this is a central point in our work.

\paragraph{Structure of the paper.} 
Section~\ref{section-preliminaries} defines the class of regular grammar 
logics
with converse via semi-Thue systems. It contains standard examples of such 
logics as well as the statement of a crucial closure property for the class of frames
of these logics.
Section~\ref{section-2gf-translation} is the core of the paper and presents
the translation into $\GF^2$ as well as obvious extensions.
In Section~\ref{section-alternative}, we present alternative
logarithmic space transformations that allow to regain the \exptime \
upper bound: one into $\GF^2$ via converse PDL with automata and
another one into the multimodal logic
with K modalities, converse, and the universal modality.
Section~\ref{section-intuitionistic-logic} illustrates how the transformation
in Section~\ref{section-2gf-translation} can be
used to translate  intuitionistic propositional logic IPL into $\GF^2$. 
Section~\ref{section-conclusion} concludes the paper and states 
open problems.

\section{Regular Grammar Logics with Converse}
\label{section-preliminaries}

Formal grammars are a convenient way of defining frame properties 
for modal logics. Many standard modal logics can be nicely defined
by a rewrite system. If one views accessibility relations as letters, 
then conditions on the accessibility relation 
can be conveniently represented by rewrite rules.
For example transitivity
 $ \forall x y z \ \relpred{\aletter}(x,y) \wedge
                   \relpred{\aletter}(y,z) \rightarrow 
                   \relpred{\aletter}(x,z) $
can be represented by the rule
$ \aletter \rightarrow \aletter \cdot \aletter. $ 
Similarly, the
inclusion
   $ \forall x y z \  \relpred{\aletter}(x,y) \wedge
                      \relpred{\aletterbis}(y,z) \rightarrow
                      \relpred{\aletterter}(x,z) $
can be represented by the rule
$ \aletterter \rightarrow \aletter \cdot \aletterbis. $
In the next 
Section~\ref{section-semi-thue-systems}, we recall 
a few definitions about formal languages, semi-Thue systems, 
and finite-state automata. 
After that, in Section~\ref{section-grammar-logics},
we introduce modal frames, and explain what it means when
a frame satisfies a rewrite rule. Using that, we can
define grammar logics. A grammar logics is {\em regular}
if the set of rewrite rules generates,
for each letter, a regular language. 
In Section~\ref{section-converse-mappings} we introduce
converse mappings. A converse mapping is a function that assigns
to each symbol $ \aletterbis $ of an alphabet a unique symbol 
$ \overline{\aletterbis}, $ {\em the converse of} $ \overline{\aletterbis}. $ 
After that we define some technical conditions on grammars
and modal frames which will ensure that converses do indeed
behave like converses. 
Section~\ref{section-converse-mappings} is concluded with an
example and a discussion of the scope of grammar logics with converse.

\subsection{Semi-Thue Systems}
\label{section-semi-thue-systems}

   An {\em alphabet} $ \aalphabet $ is a finite set 
   $ \set{\aletter_1, \ldots, \aletter_m }$ of symbols.
   We write $\aalphabet^*$ to denote
   the set of finite strings that can be
   built over elements of $ \aalphabet, $ and 
   we write $ \epsilon $ for the empty string. 
   We write $ \astring_1 \cdot \astring_2 $ for the concatenation 
   of two strings.  
   For a string $\astring \in \aalphabet^{*} $, we 
   write $ \length{\astring} $ to denote its length.
   A {\em language} over some alphabet $ \aalphabet $ 
   is defined as a subset of $ \aalphabet^* . $ 

   A {\em semi-Thue system} $\asystem$ over $\aalphabet$ is defined as 
   a subset  of $\aalphabet^* \times
   \aalphabet^*$. 
   The pairs $ ( \astring_1, \astring_2 ) \in \aalphabet^{*} \times 
                \aalphabet^{*} $ 
   are called {\em production rules}. We will mostly
   write $ \astring_1 \rightarrow \astring_2 $ instead of 
   $ (\astring_1, \astring_2) $ for production rules.
   The system $\asystem$ will be said to be {\em context-free} if
   $ \asystem $ is finite and all the production rules are in 
   $ \aalphabet \times \aalphabet^{*}. $ 
   The one-step derivation relation $ \Rightarrow_{\asystem} $ is defined 
   as follows: 
   Put $ \astring \Rightarrow_{\asystem} \astringbis $ iff there exist
   $ \astring_1, \astring_2 \in \aalphabet^{*}, $ and
   $ \astring' \rightarrow \astringbis' \in \asystem, $ such that 
   $ \astring = \astring_1 \cdot \astring' \cdot \astring_2, $ and
   $ \astringbis = \astring_1 \cdot \astringbis' \cdot \astring_2. $ 
   The full derivation relation $\Rightarrow_{\asystem}^{*} $
   is defined as the reflexive and transitive closure of 
   $ \Rightarrow_{\asystem}. $ 
   Finally, for every $\astring \in \aalphabet^*$, we 
   write $\alang_{\asystem}(\astring)$
   to denote the language $ \set{ \astringbis \in \aalphabet^{*}: 
   \astring \Rightarrow^{*}_{\asystem} \astringbis}. $

The behaviour of a context-free semi-Thue system is fully characterized by
the sets $ \alang_{\asystem}(\aletter), $ for each $ \aletter \in \aalphabet. $ 
A context-free semi-Thue system $ \asystem, $ based on 
$ \aalphabet $ is called {\em regular} if for every 
$ \aletter \in \aalphabet, $ the language $ \alang_{\asystem}(\aletter) $ is regular.
In that case,  one can associate to each $ \aletter \in \aalphabet $ 
an NDFA 
({\em non-deterministic finite automaton}, see \cite{HopcroftUllman1979})
recognizing the language $ \alang_{\asystem}(\aletter). $ 
We assume that there is some function $ \aautomaton_{\aletter}, $ that associates 
such an NDFA to each $ \aletter \in \aalphabet. $ 
We do not specify which automaton $\aautomaton_{\aletter} $ is associated
to $ \aletter. $ 
It would be possible to make $ \aautomaton_{\aletter} $ canonic, for example
by putting $\aautomaton_a $ equal to the smallest DFA recognizing
$ \alang_{\asystem}(\aletter), $ which is unique, but there seems to be no advantage in doing so. 
In many cases, an NDFA can have much less states than a corresponding DFA. 

Observe that it is undecidable to check whether 
a context-free semi-Thue system is regular since it is undecidable whether the
language generated by a linear grammar is regular (see e.g.~\cite[page 31]{Mateescu&Salomaa97}).\\

Semi-Thue systems are closely related to formal grammars, but in a 
semi-Thue system, the production rules are used for defining a 
relation between words, rather than for defining a subset of words. 
The former is precisely what we need to define grammar logics.

\subsection{Grammar Logics}
\label{section-grammar-logics}

In grammar logics, modal frame conditions are expressed by
the production rules of semi-Thue systems.
For example, transitivity on the relation $ R_{\aletter} $ is expressed
by the production rule $ \aletter \rightarrow \aletter \cdot \aletter. $ 
Similarly, reflexivity can be expressed by $ \aletter \rightarrow \epsilon. $ 

We first introduce modal languages, then we introduce modal frames and models. 
Given an alphabet $ \aalphabet, $ 
   we define the multimodal language $ {\cal L}^{\Sigma} $ based on 
   $ \aalphabet. $ In order to do this, 
   we assume a countably infinite set 
   $ \varprop = \set{\avarprop_0, \avarprop_1,\ldots} $ of 
   propositional variables.
   Then $ {\cal L}^{\Sigma} $ is recursively defined
   as follows:
   $$
   \aformula, \aformulabis ::= \avarprop \ \mid \
                               \perp \ \mid \
                               \top \ \mid \
                               \neg \aformula \ \mid \
                               \aformula \wedge \aformulabis \ \mid \
                               \aformula \vee \aformulabis \ \mid \
                               \bo{\aletter} \aformula \ \mid \
                               \diam{\aletter} \aformula           
      $$
   for $\avarprop \in \varprop$ and $\aletter \in \aalphabet. $

   We write $ \length{\aformula} $ to denote the {\em length} of the formula 
   $ \aformula $,
   that is the number of symbols needed to write $ \aformula $ down. 
   We define the {\em negation normal form} (NNF) as usual:
   $ \neg $ is applied only on members of $ \varprop. $ 
We will make use of the NNF when we translate formulas to 
$ \GF^2. $

   \label{definition-S-frames} 
   Let $ \aalphabet $ be an alphabet.  
   A {\em $ \aalphabet $-frame} is a pair 
   $ \aframe = \pair{W}{R}, $ such that
   $ W $ is non-empty, and $ R $ is a mapping from the elements of 
   $ \aalphabet $ to binary relations over $ W. $ 
   So,  for each $ \aletter \in \aalphabet, $ \ 
   $ R_{\aletter} \subseteq W \times W. $ 
   A {\em $ \aalphabet $-model} $ \amodel = \triple{W}{R}{V} $ is obtained 
   by adding a 
   valuation 
   function $ V $ with signature $ \varprop \rightarrow \powerset{W} $
   to the frame. 
   The satisfaction relation $\models$ is defined in the standard way:

   \begin{itemize}
   \itemsep 0 cm
   \item For $ \avarprop \in \varprop, \ \ \amodel, x \models \avarprop $ iff  
         $ x \in V( \avarprop ). $ 
   \item For $ \aletter \in \aalphabet, \ \ \amodel, x \models 
      \bo{\aletter} \aformula $  iff  
      for all $y, $ s.t. $ R_{\aletter}(x,y )$, \ 
      $\amodel, y \models \aformula. $
   \item For $ a \in \aalphabet, \ \ \amodel, x \models \diam{a} \aformula $ 
      iff there is an $ y, $ s.t. $ R_{a}(x,y) $ and
      $\amodel, y \models \phi. $ 
   \item
      $ \amodel, x \models \phi \wedge \psi $ iff
      $ \amodel, x \models \phi $ and 
      $ \amodel, x \models \psi. $ 
   \item
      $ \amodel, x \models \phi \vee \psi $ iff
      $ \amodel, x \models \phi $ or
      $ \amodel, x \models \psi. $
   \item
      $ \amodel, x \models \neg \phi $ iff
      it is not the case that $ \amodel, x \models \phi. $            
   \end{itemize}
   A formula $\aformula$ is said to be {\em true} 
   in the $ \aalphabet $-model $\amodel$ (written $\amodel \models \aformula$)
   iff for every $ x \in W $, \ $ \amodel, x \models \aformula $. 

   A $ \aalphabet $-frame maps the symbols in $ \Sigma $ to binary
   relations on $ W. $ This mapping can be extended to the 
   full language $ \aalphabet^{*} $ as follows:
   \begin{itemize}
   \itemsep 0 cm
   \item $ R_{\epsilon} $ equals $\set{ \pair{x}{x}: x \in W}$,
   \item For $ \astring \in \aalphabet^{*}, \aletter \in \aalphabet, \ \  
      R_{\astring \cdot \aletter} $ equals $ \set{ \pair{x}{y} : 
             \exists z \in W,  \ R_{\astring}(x,z) \mbox{ and } 
                                 R_{\aletter}(z,y) }. $ 
   \end{itemize}


\noindent
Now we define how semi-Thue systems encode conditions
on $ \aalphabet $-frames.

\begin{definition} 
   \label{definition-S-models}
   Let $ \astring \rightarrow \astringbis $ be a production rule
   over some alphabet $ \aalphabet. $ 
   We say that the $ \aalphabet $-frame $ \aframe = \pair{W}{R} $ {\em satisfies}
   $ \astring \rightarrow \astringbis $ if 
   the inclusion $ R_{\astringbis} \subseteq R_{\astring} $ holds.

   $ \aframe $ satisfies a {\em semi-Thue system} $\asystem$ if it satisfies
   each of its rules.
   We also say that $ \asystem $ {\em is true} in $ \aframe. $ 
   
   A formula $ \aformula $ is said to be 
   {\em $ \asystem $-satisfiable}  iff there is a $ \aalphabet $-model
   $\amodel = \triple{W}{R}{V} $ which satisfies $ \asystem, $ 
   and which has an $ x \in W $ such that
   $\amodel, x \models \aformula. $ 
   Similarly, a formula $ \aformula $ is said to be 
   {\em $ \asystem $-valid} iff for all $ \aalphabet $-models
   $ \amodel = \triple{W}{R}{V} $ that satisfy $ \asystem, $ 
   for all $ x \in W, $ we have 
   $ \amodel, x \models \aformula $.  
\end{definition}

We assume that a {\em logic} is characterized by its set
of satisfiable formulas, or equivalently, by its set
of universally valid formulas. 
We call logics that can be characterized by a 
semi-Thue system using Definition~\ref{definition-S-models}
{\em grammar logics}. Those logics that can be characterized
by regular semi-Thue systems are called {\em regular grammar logics}.
For instance, the modal logic S4 is the regular modal logic defined
by the regular semi-Thue system $\set{\aletter \rightarrow \epsilon, \ 
\aletter \rightarrow \aletter \aletter}$. 

\subsection{Converse Mappings}
\label{section-converse-mappings}

\noindent
In order to be able to cope with
properties such as symmetry and euclideanity, one also needs 
{\em converses}. For this reason, we 
associate to every symbol $ \aletter $ in the alphabet
a unique converse symbol $ \overline{\aletter}. $ 
Using the converse symbols, for example symmetry on the relation 
$ R_{\aletter} $ can be represented
by the production rule $ \aletter \rightarrow \overline{\aletter} $
whereas euclideanity on the relation $ R_{\aletter} $ 
can be represented
by the production rule 
$ \aletter \rightarrow \overline{\aletter} \cdot \aletter. $

\begin{definition}
   Let $ \aalphabet $ be an alphabet. We call a function
   $ \overline{\cdot} $ on $ \aalphabet $ a {\em converse mapping} if
   for all $ a \in \aalphabet, $ we have 
   $ \overline{a} \not = a $ and $ \overline{\overline{a}} = a. $ 
\end{definition}

\noindent
It is easy to prove the following: 

\begin{lemma} 
   Let $ \aalphabet $ be an alphabet with converse mapping $ \overline{\cdot} $. 
   Then $ \overline{\cdot} $ is a bijection on $ \aalphabet. $ 
   In addition, $ \aalphabet $ can be partitioned into 
   two disjoint sets $ \aalphabet^{+} $ and $ \aalphabet^{-}, $ 
   such that 
   {\bf (1)} 
      for all $ a \in \aalphabet^{+}, \ \overline{a} \in \aalphabet^{-}, $ 
   {\bf (2)} 
      for all $ a \in \aalphabet^{-}, \ \overline{a} \in \aalphabet^{+}. $ 
\end{lemma}

\noindent
In fact, there exist many partitions 
$ \aalphabet = \aalphabet^{-} \cup \aalphabet^{+}. $ 
When refer to such a partition, we assume that an arbitrary one is chosen.
We call the modal operators indexed by letters
in $\aalphabet^+$ {\em forward modalities} (conditions on successor states)
whereas 
the modal operators indexed by letters
in $ \aalphabet^{-} $ are called {\em backward modalities}
(conditions on predecessor states).

\noindent
In order to make sure that the converses really behave
like converses, we need to put the following, obvious constraint
on the $ \aalphabet $-frames. 

\begin{definition}
   \label{definition-S-converse-frames} 
   Let $ \aalphabet $ be an alphabet
   with converse mapping $ \overline{\cdot} \ . $ 
   A $ \pair{\aalphabet}{\overline{\cdot} \ } $-frame is a 
   $ \aalphabet $-frame for which in addition, 
   for each $\aletter \in \aalphabet$, 
   $R_{\overline{\aletter}} \mbox{ equals } 
      \set{ \pair{y}{x} : R_{\aletter}(x,y) }$.
\end{definition}

\begin{definition} 
   A converse mapping $ \overline{\cdot} $ can be extended to 
   words over $ \aalphabet^{*} $ as follows: 
   {\bf (1)} 
      $ \overline{\epsilon} = \epsilon, $ 
   {\bf (2)} 
      if $ u \in \Sigma^{*} $ and $ a \in \Sigma, $ then
      $ \overline{u \cdot a} = \overline{a} \cdot \overline{u}. $ 
\end{definition}


\noindent
The following property of $ \pair{\aalphabet}{\overline{\cdot}} $-frames
is easily checked:

\begin{lemma}
   \label{lemma-symmetry-extends} 
   Let $ \aalphabet $ be an alphabet with converse mapping
   $ \overline{\cdot}. $ Let $ \aframe = \pair{W}{R} $ be a
   $ \pair{\aalphabet}{\overline{\cdot} \ } $-frame.
   Then for each $ \astring \in \aalphabet^{*}$, 
   $R_{\overline{\astring}} =
             \set{ \pair{y}{x} : R_{\astring}(x,y) }$.
\end{lemma} 

%
 
\noindent
An obvious consequence of Lemma~\ref{lemma-symmetry-extends} is
that a $ \pair{\aalphabet}{\overline{\cdot} \ } $-frame
$ \pair{W}{R} $ satisfies some production rule 
$ \astring \rightarrow \astringbis $ if and only if 
$ \pair{W}{R} $ satisfies the converse
$ \overline{\astring} \rightarrow \overline{\astringbis}. $ 
Therefore, one can add the closures of the rules
to a semi-Thue system without changing the 
$ \pair{\aalphabet}{\overline{\cdot} \ } $-frames that satisfy it. 

\begin{definition}
   \label{definition-Thue-system-with-converse}
   Given a semi-Thue system $ \asystem $ over some alphabet $ \aalphabet $ 
   with converse mapping $ \overline{\cdot}, $ we call
   $ \asystem $ {\em closed under converse} if 
   $ \astring \rightarrow \astringbis \in \asystem $ implies 
   $ \overline{\astring} \rightarrow \overline{\astringbis} \in \asystem. $ 
   The {\em converse closure} of a semi-Thue system $ \asystem $ is the
   $ \subseteq $-smallest semi-Thue system $ \asystem' $ 
   closed under converse, for which $ S \subseteq \asystem'. $  
\end{definition} 

\noindent
It is easily seen that the converse closure is always well-defined. 
Because of the remark before 
Lemma~\ref{definition-Thue-system-with-converse},  
every set of 
$ \pair{\aalphabet}{\overline{\cdot} \ } $-frames 
that can be characterized by an arbitrary semi-Thue system, 
can also be characterized by a semi-Thue system which is 
closed under converse. 



We call the logics that correspond to a set 
of $ \pair{\aalphabet}{\overline{\cdot} \ } $-frames which can be
characterized by a semi-Thue system 
{\em grammar logics with converse}. 
Those logics that correspond to 
$ \pair{\aalphabet}{\overline{\cdot} \ } $-frames
which can be characterized by a regular semi-Thue system, 
are called {\em regular grammar logics with converse}.


\noindent
We give some remarks about the class of regular grammar logics with converse.

\begin{enumerate}
\itemsep 0 cm
\item 
   Because of the remarks before
   Definition~\ref{definition-Thue-system-with-converse}, 
   allowing or disallowing semi-Thue systems 
   which are not closed under converse  
   does not have consequences for the logics one can define. 
\item 
   It can be easily checked that every regular grammar logic 
   (without converse) can be viewed as a fragment of 
   a regular grammar logic with converse.
   This is due to the fact that one can add a set of converses 
   $ \overline{\aalphabet} $ to the alphabet 
   $ \aalphabet $ and extend the $ \aalphabet $-frame into a 
   $ \pair{\aalphabet \cup \overline{\aalphabet}}{\overline{\cdot} \ } $-frame.

\item 
   Originally, grammar logics were defined with formal grammars in
   \cite{Farinas&Penttonen88} (as in~\cite{Baldoni98,Demri01,Demri01b}), and 
   they form a subclass of Sahlqvist modal logics~\cite{Sahlqvist75} with 
   frame conditions
   expressible in $\Pi_1$ when $\asystem$ is context-free.  
   $\Pi_1$ is the class of first-order formulae of the form
   $\forall \ x_1 \ \forall \ x_2 \ \ldots \forall \ x_n \ \aformula$ where
   $\aformula$ is quantifier-free. 
   In the present paper, we adopt a lighter presentation based on 
   semi-Thue systems as done in~\cite{Chagrov&Shehtman94},
   which is more appropriate. 

\end{enumerate}

\begin{example}
   The standard modal logics K, T, B, S4, K5, K45, and S5 can be 
   defined as regular grammar logics 
   over the singleton alphabet $ \aalphabet = \{ \aletter \}. $ 
   In Table~\ref{table-st}, we specify the semi-Thue systems through 
   regular expressions for the languages $ \alang_{\asystem}(\aletter). $ 

\begin{table}
\begin{center}
 \begin{tabular}{|c|c|c|}
 \hline \hline
  logic  & $\alang_{\asystem}(\aletter)$ & frame condition \\ \hline
K & $\set{\aletter}$ & (none) \\ \hline
KT & $\set{\aletter, \epsilon}$ & reflexivity \\ \hline
KB & $ \set{\aletter, \overline{\aletter}} $ & symmetry \\ \hline
KTB & $ \set{\aletter, \overline{\aletter}, \epsilon} $ & refl. and sym. \\ \hline
K4 & $\set{\aletter} \cdot \set{\aletter}^* $ & transitivity \\ \hline
KT4 = S4 & $\set{\aletter}^*$ & refl. and trans. \\ \hline
KB4 & $\set{\aletter, \overline{\aletter}} \cdot \set{\aletter, 
\overline{\aletter}}^*$ & sym. and trans. \\ \hline 
K5 & $(\set{ \overline{\aletter}} \cdot \set{\aletter, \overline{\aletter}}^* \cdot 
\set{\aletter})
 \cup \set{\aletter}$  & euclideanity \\ \hline
KT5 = S5 & $\set{\aletter, \overline{\aletter}}^*$ & equivalence rel. \\ \hline
K45 & $(\set{ \overline{\aletter}}^* \cdot \set{\aletter})^{*}$ & trans. and eucl. \\
\hline
\hline
 \end{tabular}
 \end{center}
\caption{Regular languages for standard modal logics}
\label{table-st}
\end{table}
\end{example}

\noindent
Numerous other logics for specific application domains are in fact regular
grammar logics with converse, or logics that can be reduced to such
logics. We list below some examples:

\begin{itemize}
\item 
   description logics (with role hierarchy, transitive roles), 
   see e.g.~\cite{Horrocks&Sattler99};

\item 
   knowledge logics, see e.g. S5$_m$(DE) in
   ~\cite{Fagin&Halpern&Moses&Vardi95};

\item 
   bimodal logics for intuitionistic modal logics of the form {\bf IntK}$_{\Box} \ +
  \Gamma$~\cite{Wolter&Zakharyaschev97}. Indeed, let $\asystem$ be a regular
semi-Thue system (over $ \aalphabet $) closed under converse and let
$\aalphabet' \subset
\aalphabet$ be such that for every $\aletter \in \aalphabet$, either
$\aletter \not \in \aalphabet'$ or $\overline{\aletter} \not \in \aalphabet'$.
Then, the semi-Thue system $\asystem \cup 
\set{\aletterbis \rightarrow \aletterbis \aletter \aletterbis, 
     \overline{\aletterbis} \rightarrow \overline{\aletterbis} 
     \overline{\aletter} \overline{\aletterbis}: \aletter \in \aalphabet'}$
over $\aalphabet \cup \set{\aletterbis, \overline{\aletterbis}}$ 
is also regular, assuming $\aletterbis, \overline{\aletterbis} \not \in \aalphabet$. 
By taking advantage of~\cite{Ganzinger&Meyer&Veanes99},
in ~\cite{Alechina&Shkatov03} decidability of intuitionistic modal logics is 
also shown in a uniform manner.

\item fragments of logics designed for the access control in distributed 
      systems \cite{Abadietal93,Massacci97}.

\item 
   extensions with the universal modality~\cite{Goranko&Passy92}.
   Indeed, for every regular grammar logic with converse,
   its extension with a universal modal operator  is also a 
   regular grammar logic with converse by using simple arguments from
   \cite{Goranko&Passy92} (add an S5 modal connective stronger than any other
   modal connective).
   Hence, satisfiability, global satisfiability and logical consequence can be
   handled uniformly with no increase of worst-case complexity;
\item 
   information logics, see e.g. \cite{Vakarelov87}.
   For instance, the 
  Nondeterministic Information Logic NIL introduced in~\cite{Vakarelov87}
   (see also~\cite{Demri00c})
  can be shown to be  a fragment of a regular grammar logic with converse with $\aalphabet^+
  = \set{fin,sim}$ and the converse closure of the production rules below:
  \begin{itemize}
  \itemsep 0 cm
  \item $fin \rightarrow fin \cdot  fin$; $fin \rightarrow \epsilon$;
  \item $sim \rightarrow \overline{sim}$; $sim \rightarrow \epsilon$;
  \item $sim \rightarrow \overline{fin} \cdot sim \cdot fin$.
  \end{itemize}
  For instance $\alang_{\asystem}(sim) = \set{\overline{fin}}^* \cdot
  \set{sim, \overline{sim}, \epsilon} \cdot \set{fin}^*$.                      
\end{itemize}

\paragraph{A frame condition outside our current framework.}
The euclideanity condition can be slightly generalized by considering
frame conditions of the form \\
$(R_{\aletter}^{-1})^{n} ; R_{\aletter} \subseteq R_{\aletter}$ for some $n \geq 1$.
The context-free semi-Thue system corresponding
to this inclusion is $\asystem_n = \set{
\aletter \rightarrow \overline{\aletter}^n \aletter, \ 
\overline{\aletter} \rightarrow \overline{\aletter} \aletter^n }$.
The case $n = 1$ corresponds to euclideanity.
Although we have seen that for $n = 1$, the language 
$\alang_{\asystem_1}(\aletter)$ is regular, one can establish that
in general, for $ n > 1, $ the language 
$\alang_{\asystem_n}(\aletter)$ is not regular. 
This is particularly interesting since $\asystem_n$-satisfiability
restricted to formulae with only the modal operator $\bo{\aletter}$ is 
decidable, see e.g.~\cite{Gabbay75b,Hustadt&Schmidt03}.
To see why the languages $\alang_{\asystem_n}(\aletter)$ are not regular, 
consider strings of the following form:
\[ \sigma_n(i_1,i_2) = ( \overline{\aletter} \aletter^{n-1} )^{i_1} \ a \ 
                     ( \overline{\aletter}^{n-1} a )^{i_2}. \]
\[ \overline{\sigma}_n(i_1,i_2) =
                ( \overline{\aletter} \aletter^{n-1} )^{i_1} \ \overline{a} \
                ( \overline{\aletter}^{n-1} a )^{i_2}. \]
We show that 
\[ ( \ a \Rightarrow^{*}_{\asystem_n} \sigma_n(i_1, i_2) \mbox{ and } 
       a \Rightarrow^{*}_{\asystem_n} \overline{\sigma}_n(i_1,i_2 + 1) \ ) 
              \mbox{ iff } 
      i_1 = i_2. \]
In order to check that the equivalence holds from right to left,
observe that $ a = \sigma_n(0,0), $ and 
\[ \sigma_n(0,0) \Rightarrow_{\asystem_n}
   \overline{\sigma}_n(0,1) \Rightarrow_{\asystem_n} 
   \sigma_n(1,1) \Rightarrow_{\asystem_n} \cdots \]
\[ \Rightarrow_{\asystem_n} 
   \sigma_n(i,i) \Rightarrow_{\asystem_n} 
   \overline{\sigma}_n(i,i+1) \Rightarrow_{\asystem_n}
   \sigma_n(i+1,i+1) \Rightarrow_{\asystem_n} \cdots \]

\noindent
We now prove the equivalence from left to right. Let us say
that $ \astring $ is {\em an predecessor of} $ \astringbis $ if
$ \astring \Rightarrow_{{\asystem}_n} \astringbis. $ 
Then it is sufficient to observe the following: 
\begin{enumerate}
\item
   A string of form $ \sigma_n(0,j) $ with $j > 0$ has no predecessor. 
\item
   A string of form $ \sigma_n(i+1,j) $ has only one predecessor, namely
   $ \overline{\sigma}_n(i,j). $ 
\item
   A string of form $ \overline{\sigma}_n(i,0) $ has no predecessor.
\item
   A string of form $ \overline{\sigma}_n(i,j+1) $ has only one predecessor,
   namely $ \sigma_n(i,j). $ 
\end{enumerate}
To have a predecessor, a string must have a sequence of at least
$ n $ consecutive $ \aletter $'s or $ \overline{\aletter} $'s. 
The strings of form $ 1 $ or $ 3 $ have no such sequence. The strings
of form $ 2 $ or $ 4 $ have exactly one such sequence. 

\noindent
We have
$$ \alang_{\asystem_n}(\aletter) \cap \set{ \sigma_n(i,j) : i \geq 0, \ j
  \geq 0 } =
   \set{ \sigma_n(i,i) : i \geq 0 }.
$$ 
$\set{ \sigma_n(i,i) : i \geq 0 }$ is clearly not regular (we assume $n > 1$)
and $\set{ \sigma_n(i,j) : i \geq 0, \ j \geq 0 }$ is clearly regular.
Since the regular languages are closed under intersection, 
$\alang_{\asystem_n}(\aletter)$ cannot be regular for $n > 1$.
%
%
Hence, this will leave open the extension of our translation method  to
the case of context-free semi-Thue systems with converse when
decidability holds (see e.g. decidable extensions of PDL with certain
context-free programs in~\cite{Harel&Kozen&Tiuryn00}).

\subsection{Characterizations of 
            $ \pair{\aalphabet}{\overline{\cdot} \ } $-Frames}

Theorem \ref{theorem-three-characterizations} below states
the usual relations between derivations, validity and frame conditions.

\begin{theorem}
   \label{theorem-three-characterizations}
   Let $ \aalphabet $ be an alphabet, $ \astring, \astringbis \in
   \aalphabet^{*}$, and
   $ \asystem $ be a context-free semi-Thue system over alphabet
   $ \aalphabet, $ which is closed under converse. 
   Consider the statements below: 
   \begin{description}
   \item[(I)] $\astring \Rightarrow_{\asystem}^{*} \astringbis$. 
   \item[(II)] In every $ \pair{\aalphabet}{\overline{\cdot} \ } $-frame
               $ \aframe $ satisfying $ \asystem, $ 
               for arbitrary $ p \in \varprop, $ \\ 
                $\bo{\astring} \avarprop \Rightarrow 
               \bo{\astringbis} \avarprop$ is valid. \\
               (For a word $ \astring = (\astring_1,\ldots, \astring_m)$, 
               $ \bo{\astring} \avarprop $ is an
               abbreviation for
               $ \bo{\astring_1} \cdots \bo{\astring_{m}} \avarprop $.)
   \item[(III)] $ R_{\astringbis} \subseteq R_{\astring}$
                in every $
               \pair{\aalphabet}{\overline{\cdot} \ } $-frame $\aframe$
               satisfying $\asystem$.\\
                (This is the same as saying that $ \aframe $ makes
                  $ \astring \rightarrow \astringbis $ true.)
   \end{description}
Then, (II) is equivalent to (III), (I) implies (II), but (II) does not
necessarily imply (I).
\end{theorem}

\noindent
The equivalence between (II) and (III) is a classical correspondence
result in modal logic theory (see e.g.,~\cite{vanBenthem84c}).
The proof does not make use of the fact that the frame $ \aframe $ is a
$ \pair{\aalphabet}{\overline{\cdot} \ } $-frame. 
(I) implies (III) holds for every $ \aalphabet $-frame, and therefore
also for every $ \pair{\aalphabet}{\overline{\cdot} \ } $-frame.
It can be proved by induction on the length
of the derivation of 
$ \astring \Rightarrow_{\asystem}^{*} \astringbis. $
More precisely, when (I) holds, then there is an $ i, $ such that
$ \astring \Rightarrow_{\asystem}^{i} \astringbis. $ 
Then (III) is proven by induction on $i$.

In order to show that (II) does not necessarily imply 
(I), consider the
   semi-Thue system $ \asystem = 
      \set{ \aletter \rightarrow \overline{\aletter}, \ 
            \overline{\aletter} \rightarrow \aletter, \ 
            \aletterbis \rightarrow \aletter^{3}, \ 
            \overline{\aletterbis}\rightarrow 
                                       \overline{\aletter}^3 }$. 
   In this system, $ \aletterbis \not \Rightarrow^*_{\asystem} \aletter. $ 
   However, $ \aletter \rightarrow \overline{\aletter} $ expresses
   symmetry, which in a $ \pair{\aalphabet}{\overline{\cdot} \ } $-frame
   satisfying $\asystem$
   implies that whenever $ \pair{x}{y} \in R_{\aletter}, $ then
   also $ \pair{x}{y} \in R_{\aletter^3}. $ 
   Therefore,  
   $ R_{\aletter} \subseteq R_{\aletterbis}$.

It is worth observing that in the absence of converse, (II) implies
(I) by the proof of~\cite[Theorem 3]{Chagrov&Shehtman94} (see also the
tableaux-based
proof in~\cite{Baldoni98}).
This is based on the fact that every ordered monoid is embeddable into some
ordered monoid of binary relations (see more details in 
\cite{Chagrov&Shehtman94}).
In the presence of converse, this property does not hold since 
it is not true that
every tense ordered monoid is embeddable into some
tense ordered monoid of binary relations.

In Section~\ref{correctness-of-translation}, we need only the 
implication (I) $ \Rightarrow $ (III) 
for proving the first direction of Theorem~\ref{theorem-main-translation}.

If some $ \aalphabet $-frame $ \aframe = \pair{W}{R} $ does  satisfy a 
regular semi-Thue system, then one can add missing edges 
to $ R $ until the resulting $ \aalphabet $-frame does satisfy the 
semi-Thue system.
In this way, one obtains a closure function that assigns to 
each $ \aalphabet $-frame the smallest frame that satisfies 
the semi-Thue system.

\begin{definition} 
   \label{definition-closure-operator} 
   We first define an inclusion relation on frames.
   Let $ \aalphabet $ be an alphabet. Let
   $ \aframe_1 = \pair{W}{R_1} $ and 
   $ \aframe_2 = \pair{W}{R_2} $ be two $ \aalphabet $-frames 
   with the same set of worlds $ W. $ 
   We say that $ \aframe_1 $ is a {\em subframe} of 
               $ \aframe_2 $ if 
   for every $ \aletter \in \aalphabet, $ \ \ 
   $ R_{1,\aletter} \subseteq R_{2,\aletter}. $

   Using this, we define the {\em closure operator} $ C_{\asystem} $
   as follows: 
   For every context-free semi-Thue system $ \asystem $ over
   alphabet $ \aalphabet, $ 
   for every $ \aalphabet $-frame $ \aframe, $ 
   the {\em closure} of $ \aframe $ under $ \asystem $ is defined
   as the smallest $\aalphabet$-frame (under the subframe relation) that
   satisfies $ \asystem, $ and which has $ \aframe $ as 
   a subframe. We write $ C_{\asystem}(\aframe) $ for the
   closure of $ \aframe. $ 
\end{definition} 

The closure always exists, and 
is unique, because of the Knaster-Tarski fixed point theorem.
It can also be proven from the forthcoming
Theorem~\ref{theorem-closure-is-lazy}.
Definition~\ref{definition-closure-operator} does not mention
converse mappings. We will later
(Lemma~\ref{lemma-closure-converse}) show that $ C_{\asystem} $ transforms
$ \pair{\aalphabet}{\overline{\cdot} \ } $-frames into
$ \pair{\aalphabet}{\overline{\cdot} \ } $-frames,
even if $ \asystem $ is not closed under converse. 

%

We now prove a crucial property of $ C_{\asystem}, $ 
namely that every edge added by $ C_{\asystem} $ can be justified
in terms of an $ \alang_{\asystem}(\aletter). $ 

\begin{theorem} 
   \label{theorem-closure-is-lazy}
   Let $ \asystem $ be a context-free semi-Thue
   system, which is closed under converse.
   Let $ \aframe = \pair{W}{R} $ be a $\aalphabet$-frame. 
   Let the $\aalphabet$-frame $ \aframe' = \pair{W}{R'} $ be defined from
   \[ R'_{a} = 
      \bigcup_{\astring \in \alang_{\asystem}(\aletter)} R_{\astring},
      \mbox{ for } a \in \aalphabet. \]
   Then $ \aframe' = C_{\asystem}( \aframe ). $ 
\end{theorem}

\begin{proof} 
   We have to show that
   \begin{enumerate}
   \itemsep 0 cm
   \item
      $ \pair{W}{R'} $ satisfies $ \asystem, $ and
   \item
      among the frames that satisfy $ \asystem, $ and that 
      have $ \aframe $ as subframe, $ \pair{W}{R'} $ is a 
      minimal such frame.
   \end{enumerate}
   In order to show {\bf (1)}, we show that for every rule
   $ \aletter \rightarrow \astring $ in $ \asystem, $ the inclusion 
   $ R'_{\astring} \subseteq R'_{\aletter} $ holds. 
   Write $ u = (\astring_1, \ldots, \astring_n), $ with $ n \geq 0, $ and
   each $ u_i \in \aalphabet. $ 
   Let $ \pair{x}{y} \in R'_{\astring}. $ We intend to show that
   $ \pair{x}{y} \in R'_{\aletter}. $ 

   \noindent
   By definition, there are $ z_1, \ldots, z_{n-1} \in W, $ such that
   \[ \pair{x}{z_1} \in R'_{\astring_1}, \ \ 
      \pair{z_1}{z_2} \in R'_{\astring_2}, \ldots, \ \ 
      \pair{z_{n-1}}{y} \in R'_{\astring_n}. \]
   By construction of $ R', $ there are words 
   $ \astringbis_1, \ldots, \astringbis_n \in \aalphabet^{*}, $ such that \\ 
   $ u_1 \Rightarrow^{*}_{\asystem} \astringbis_1, \ \ \ldots, \ \ 
     u_n \Rightarrow^{*}_{\asystem} \astringbis_n, $ and
   \[ \pair{x}{z_1} \in R_{\astringbis_1}, \ \ 
      \pair{z_1}{z_2} \in R_{\astringbis_2}, \ldots, \ \ 
      \pair{z_{n-1}}{y} \in R_{\astringbis_n}. \]
   As a consequence,
   $ \pair{x}{y} \in R_{\astringbis_1 \cdot  \ldots \cdot \astringbis_n} $  
   Now because $ a \Rightarrow_{\asystem} \astring, $ \ \ 
   $ \astring = ( \astring_1, \ldots, \astring_n ), $ and
   each $ \astring_i \Rightarrow^{*}_{\asystem} \astringbis_i, $ we
   also have $ \aletter \Rightarrow^{*}_{\asystem} 
   \astringbis_1 \cdot \ldots \cdot \astringbis_n. $ 
   It follows that 
   $ \pair{x}{y} \in R'_{a}, $ by the way $ R' $ is constructed.

   \noindent
   Next we show {\bf (2)}.
   Let $ \pair{W}{R''} $ be a $\aalphabet$-frame, such that $ \pair{W}{R} $ 
   is a subframe of $ \pair{W}{R''} $ and 
   $ \pair{W}{R''} $ satisfies $ \asystem. $ 
   We want to show that for each $ \aletter \in \aalphabet, $ \ \ 
   $ R'_{\aletter} \subseteq R''_{\aletter}. $ 

   If $ \pair{x}{y} \in R'_{\aletter}, $ this means that 
   $ \pair{x}{y} \in R_{\astring} $ for a $ \astring $ with 
   $ \aletter \Rightarrow^{*}_{\asystem} \astring. $ 
   Because $ R_{\astring} \subseteq R''_{\astring}, $ we also have
   $ \pair{x}{y} \in R''_{\astring}. $ 
   Because $ \pair{W}{R''} $ is an $ \aalphabet$-frame satisfying 
   $\asystem$, it follows
   from Theorem~\ref{theorem-three-characterizations} 
   ( (I) $ \Rightarrow $ (III) ) that
   $ R''_{\astring} \subseteq R''_{\aletter}, $ 
   and we have $ \pair{x}{y} \in R''_{\aletter}. $ 
\end{proof}

\begin{lemma} \label{lemma-closure-converse}
   Let $ \aalphabet $ be an alphabet with converse mapping $ \overline{\cdot}
   \ $ and let $ \aframe = \pair{W}{R} $ be a 
   $ \pair{\aalphabet}{\overline{\cdot} \ } $-frame.
   Then for every context-free semi-Thue system $ \asystem $ 
   over $ \aalphabet, $ 
   the closure $ C_{\asystem}( \aframe ) $ is also a 
   $ \pair{\aalphabet}{\overline{\cdot} \ } $-frame.
\end{lemma} 

\begin{proof}
   Write $ C_{\asystem} ( \aframe ) = \pair{W}{R'}. $ 
   We need to show that for each $ \aletter \in \aalphabet, $ 
   \[ R'_{\aletter} = \set{ \pair{x}{y} \ | \ 
                            \pair{y}{x} \in R'_{\overline{a}} }. \]

   \noindent
   In case that this not hold, one can define $ R'' $ from
   \[ R''_{\aletter} = \set{ \pair{x}{y} \ | \ 
                         \pair{x}{y} \in R'_{\aletter} \mbox{ and }
                         \pair{y}{x} \in R'_{\overline{\aletter}} },
                         \mbox{ for } \aletter \in \aalphabet. \]
   and $ \pair{W}{R''} $ is a strict subframe of $ \pair{W}{R'} $ 
   satisfying $ \asystem. $ 
   Because also $ \pair{W}{R} $ is a subframe of $ \pair{W}{R''}, $ 
   we obtained a contradiction with the minimality of $ \pair{W}{R}. $ 

%
\end{proof}

When $\asystem$ is regular, the map $C_{\asystem}$ is a monadic second-order definable graph transduction in the sense of~\cite{Courcelle94} and it is
precisely the inverse substitution $h^{-1}$ in the sense of~\cite{Caucal03}
(see also~\cite{Caucal96}) when the extended substitution $h$ is defined
by $\aletter \in \aalphabet \mapsto \alang_{\asystem}(\aletter)$.

\section{The Translation into $\GF^2$}
\label{section-2gf-translation}

In this chapter, we define the transformation from regular
grammar logics with converse into $ \GF^2. $ 
The transformation can be carried out in logarithmic space. 
When translating a $ \Box $-subformula, the
translation simulates the behaviour of an NDFA
in order to determine to which worlds the $ \Box $-formula
applies. 
This generalises the results in
\cite{deNivelle99,deNivelle01} for the logics S4 and K5,
which were at an ad hoc basis. Here we
show that it is the regularity of the frame condition that makes
the translation method work.
The translation allows us to provide an \exptime \ upper bound
for the satisfiability
problem for regular grammar logics with converse.
Other specific features of our translation are the following ones:
\begin{itemize}
\itemsep 0 cm
\item This is not an exact translation of the Kripke semantics  
   of modal logics, which makes it different from the relational 
   translation. Indeed, we rather
   define a transformation from the satisfiability problem for
   a regular grammar logic with converse into the satisfiability problem
   for $ \GF^2$. Hence, our translation is a reduction as understood in
   complexity theory, see e.g.~\cite{Papadimitriou94}.
\item 
   The translation is based on a mutual recursion between the encoding of
   the frame conditions and the translation of logical operators.
\end{itemize}

\subsection{The Transformation}
\label{section-transformation}

   We assume that $ \asystem $ is a regular semi-Thue system 
   closed under converse over 
   an alphabet $ \aalphabet $ with converse mapping $ \overline{\cdot}$
   (and partition $\set{\aalphabet^+, \aalphabet^-}$). 
   For every $ \aletter \in \aalphabet, $ the automaton
   $ \aautomaton_{\aletter} $ is an NDFA (possibly with $\epsilon$-transitions) 
   recognizing the language $ \alang_{\asystem}(\aletter). $  
   It is in principle allowed that $ \aautomaton_{\aletter} $ and
   $ \aautomaton_{\overline{\aletter}} $ are unrelated automata, 
   although they have to accept isomorphic languages
   (because $\astring \in \alang_{\asystem}(\aletter)$ iff 
    $\overline{\astring} \in \alang_{\asystem}(\overline{\aletter})$ for every
   $\astring \in \aalphabet^*$).
   We write $ \aautomaton_{\aletter} = 
      ( Q_{\aletter}, s_{\aletter}, F_{\aletter}, \delta_{\aletter} ). $
   Here $ Q_{\aletter} $ is the finite set of states, $ s_{\aletter} $ 
   is the starting state,
   $ F_{\aletter} \subseteq Q_{\aletter} $ is the set of accepting states, 
   and $ \delta_{\aletter} $ is the transition function,
   which is possibly non-deterministic. 
   When all rules in $\asystem$ are either 
   right-linear\footnote{i.e., there is a partition 
   $\set{V,T}$ of $\aalphabet$
   such that the production rules are in  $V \rightarrow T^* \cdot (V
   \cup \set{\epsilon})$.} or left-linear\footnote{i.e., there is a partition $\set{V,T}$ of $\aalphabet$
   such that the production rules are in  $V \rightarrow (V \cup \set{\epsilon}) \cdot
T^*$.}, then each automaton 
   $\aautomaton_{\aletter}$ can be effectively built in logarithmic space in
   $\length{\asystem}$, the size of $\asystem$ with some reasonably succinct encoding. 

In the sequel we assume that the two variables in $\GF^2$ are $\set{x_0,x_1}$.
$\alpha$ and $\beta$ are used as distinct meta-variables in $\set{x_0,x_1}$.
Observe that in Definition~\ref{translation-formula} 
the quantification alternates
over $\alpha$ and $\beta$.

   \begin{definition} 
      Assume that for each letter $ \aletter \in \aalphabet^{+}, $ 
      a unique binary predicate symbol $ {\bf R}_{\aletter} $ is given.
      We define a translation function $ t_{\aletter}, $ mapping
      letters in $ \Sigma $ to binary predicates. 
      \begin{itemize}
      \itemsep 0 cm
      \item      
         For each letter $ \aletter \in \aalphabet^{+}, $ 
         we define $ t_{\aletter}(\alpha,\beta) = 
                        {\bf R}_{\aletter}(\alpha,\beta), $ 
      \item
         For each letter $ \aletter \in \aalphabet^{-}, $ 
         we define 
         $ t_{\aletter}(\alpha,\beta) = 
                        {\bf R}_{\aletter}(\beta,\alpha). $ 
      \end{itemize} 
   \end{definition} 

   We could have separated the symmetry from the function $ t $ itself,
   in the same way as we did with frames in
   Definition~\ref{definition-S-frames} and 
   Definition~\ref{definition-S-converse-frames}.
   Then we would have first defined $ t_{\aletter} $ for each
   $ \aletter \in \aalphabet $ without any conditions, and 
   later defined that some $ t $ {\em respects} $ \cdot $ 
   if always $ t_{\aletter} $ is the converse of 
   $ t_{\overline{\aletter}}. $ 
   This would in principle be elegant and allow us to specify
   more clearly which property of the translation depends on which 
   part of the definition of the translation. However it would
   make the formulations of the properties more tedious, so 
   we decided not to do this. 

   We now define the main part of the translation.
   It takes two parameters, a one-place first-order formula and an 
   NDFA. The result of the translation is a first-order formula
   (one-place again) that has the following meaning:
   \begin{quote}
      In every point that
      is reachable by a sequence of transitions that are accepted
      by the automaton, the original one-place formula holds. 
   \end{quote} 
    
   \begin{definition}
      \label{definition-translation-automaton} 
      Let $ \aautomaton =  \triple{Q}{s, F}{\delta}$ be an NDFA.
      Let $ \aformulater(\alpha) $ be a {\em first-order} formula with
      one free variable $ \alpha$. 
      Assume that for each state $\astate \in Q$,  a 
      fresh unary predicate symbol $ {\bf q} $ is given. 
      We define $ t_{\aautomaton}( \alpha, \aformulater ) $ as the 
      conjunction of the following formulas (the purpose of the first 
      argument is to remember that $ \alpha $ is the free variable of 
      $ \aformulater $): 
      \begin{itemize}
      \itemsep 0 cm
      \item
         For the initial state $s$, the formula
         $ {\bf s}(\alpha) $ is included in the conjunction.
      \item
         For each $\astate \in Q$, for each $ \aletter \in \aalphabet$, 
         for each $ r \in \delta(\astate, \aletter)$, the formula
         \[ \forall \alpha \beta \ 
              [ \ t_{\aletter}(\alpha,\beta) \rightarrow 
                {\bf q}(\alpha) \rightarrow 
                {\bf r}(\beta) \ ] \] 
         is included in the conjunction. 
      \item
         For each $ \astate \in Q, $ for each 
         $ r \in \delta(\astate, \epsilon), $ the formula 
         \[ \forall \alpha \ 
               [ \ {\bf q}(\alpha) \rightarrow 
                   {\bf r}(\alpha) \ ] \]
         is included in the conjunction. 
      \item
         For each $ \astate \in F, $ the formula 
         \[ \forall \alpha \ [ \ {\bf \astate}(\alpha) \rightarrow
                                 \aformulater(\alpha) \ ] \]
         is included in the conjunction. 
      \end{itemize} 
   \end{definition} 


The function $ t_{\aautomaton}( \alpha, \aformulabis ) $ is
applied on formulas $ \aformulabis $ that are subformulas
of an initial formula $ \aformula. $ 
The definition requires that in each
application of $ t_{\aautomaton}, $
distinct predicate symbols of form $ {\bf q} $ for $ q \in Q $ 
are introduced. This can be done 
either occurrence-wise, or subformula-wise.
Occurrence-wise means that, if some subformula $ \aformulabis $ of
$ \aformula $ occurs more than once, then different fresh predicate
symbols have to be introduced for each occurrence.
Subformula-wise means that the different occurrences can share
the fresh predicates.
In the sequel, we will assume the subformula-wise approach.

If the automaton $ \aautomaton $ has more than one accepting
state, then $ \aformulater(\alpha) $ occurs more than
once in the translation $ t_{\aautomaton}( \alpha, \aformulater ) $. 
This may cause an exponential blow-up in the translation process but
this problem can be easily
solved by adding a new accepting state to the automaton,
and adding $ \epsilon $-translations from the old accepting
states into the new accepting state. 

Now we can give the translation itself.
It behaves like a standard relational translation on all subformulas,
except for those of the form $ \bo{\aletter} \aformulabis$, on which 
$ t_{\aautomaton_{\aletter}} $ will be used.
In order to easily recognize the $ \Box $-subformulas,
we require the formula $ \aformula $ to be in negation 
normal form.
One could define the translation without it,
but it would have more cases. 

\begin{definition} 
   \label{translation-formula} 
   Let $ \aformula \in {\cal L}^{\aalphabet} $ be 
   a modal formula in NNF. Let $ \asystem $ be a regular
   semi-Thue system closed under converse over alphabet $ \aalphabet $ with
   converse mapping $ \overline{\cdot}. $ 
   Assume that for each $ \aletter \in \aalphabet $ an 
   automaton $ \aautomaton_{\aletter} $ recognizing
   $ \alang_{\asystem}( \aletter ) $ is given.
   We define the translation $ T_{\asystem}(\aformula) $
   as $ t( \aformula, x_0, x_1) $
   from the following
   function $ t( \aformulabis, \alpha, \beta )$, which
   is defined 
   by recursion on the subformulas $ \aformulabis $ of 
   $ \aformula:$ 
   \begin{itemize}
   \itemsep 0 cm
   \item 
      $ t(\avarprop, \alpha, \beta) $ equals $ \proppred{p}(\alpha), $ 
      where $\proppred{p}$ is
      a unary predicate symbol uniquely associated to the propositional
      variable $\avarprop$. 
   \item $t(\neg \avarprop, \alpha, \beta) $ equals 
      $ \neg \proppred{p}(\alpha), $
   \item $t(\aformulabis \wedge \aformulabis', \alpha, \beta) $ equals 
      $ t(\aformulabis, \alpha, \beta) \wedge t(\aformulabis', \alpha, \beta), $ 
   \item $t(\aformulabis \vee \aformulabis', \alpha, \beta) $ equals
      $ t(\aformulabis, \alpha, \beta) \vee t(\aformulabis', \alpha, \beta), $ 
   \item 
      for $ a \in \aalphabet, \ \  
      t(\diam{a} \aformulabis, \ \alpha, \beta) $ equals 
      $ \exists \beta \ [ \ t_{\aletter}(\alpha, \beta) \wedge 
      t(\aformulabis, \beta, \alpha) \ ], $ 
   \item 
      for $ \aletter \in \aalphabet, \ \ 
        t(\bo{\aletter} \aformulabis, \ \alpha, \beta ) $ equals 
      $ t_{\aautomaton_{\aletter}} (\alpha, t(\aformulabis, \alpha, \beta) \ ). $ 
   \end{itemize} 
\end{definition}

When translating a subformula of form $ \bo{\aletter} \aformulabis, $ 
the translation function 
$ t_{\aautomaton_{\aletter}} $ of Definition~\ref{definition-translation-automaton}
is used. 
The only difference with the standard relational translation is the
translation of $ \bo{\aletter} $-formulae.

\begin{lemma}
   For the translation $T_{\asystem}(\aformula)$, the following holds:
   \begin{description}
   \itemsep 0 cm
   \item[(I)] The only variables occurring in $ T_{\asystem}(\aformula) $ 
      are in $\set{x_0,x_1}$ and $\alpha$ is the only free variable in $t(\aformulabis, \alpha, \beta)$.      
   \item[(II)] $ T_{\asystem}(\aformula) $ is in the guarded fragment.
   \item[(III)]
      The size of $T_{\asystem}(\aformula)$ is in 
      ${\cal O}(\length{\aformula} \times m)$.
   \item[(IV)]
      $ T_{\asystem}(\aformula ) $ can be computed in logarithmic space
      in $\length{\aformula} + m. $
   \end{description} 
   Here $ m $ is the size of the largest $ \aautomaton_{\aletter}, $ 
   i.e. 
   $ m = {\rm max} 
         \set{ \ \length{\aautomaton_{\aletter}} \ \ | \ \   
               \aletter \in \aalphabet \ }. $ 
\end{lemma}

When $\asystem$ is formed of production rules of a formal grammar that is
   either right-linear or left-linear, then
   $m$ is in $\mathcal{O}(\length{\asystem})$.
For a given semi-Thue system $ \asystem, $ the number $ m $ is
fixed. As a consequence, $ T_{\asystem}(\aformula) $ has size 
linear in $ \length{\aformula} $ for a given logic. 

Unlike the standard relational translation from modal logic into classical
predicate
logic (see e.g., \cite{Fine75b,vanBenthem76,Morgan76,Moore77}),
the 
subformulae in $T_{\asystem}(\aformula)$
mix the frame conditions and the interpretation of the logical
connectives.
Besides $ T_{\asystem}(\aformula) $ can be viewed as the logical counterpart of the
propagation rules defined in~\cite{Castilho&Farinas&Gasquet&Herzig97} 
(see also~\cite{Gore99,Massacci00,Farinas&Gasquet02}).


\begin{example} 
   \label{example-k5}
   Let $\aformula = \Diamond \avarprop \wedge \Diamond \Box \neg \avarprop$ 
   be the negation normal form of the formula
   $\neg (\Diamond \avarprop \Rightarrow \Box \Diamond
   \avarprop)$.
   We consider K5, and assume one modality $ \aletter, $  
   so $ \Box $ is an abbreviation for $ \bo{\aletter}, $ and
   $ \Diamond $ is an abbreviation for $ \diam{\aletter}. $ 
   Table~\ref{table-k5} contains to the left an automaton 
   $ {\cal A}_{\aletter} $ recognizing
   the language defined in Table~\ref{table-st} for K5. 
   To the right is the translation 
   $ t_{\aautomaton_{\aletter}}( \alpha, \aformulater(\alpha))$ 
   for some first-order formula $ \aformulater(\alpha). $ 
   The translation $T_{\asystem}(\aformula)$ of $ \aformula $ is equal to
   \[
      \exists \beta \ [ \ \relpred{}(\alpha,\beta) \wedge 
                          \proppred{p}(\beta) \ ]
      \wedge
      \exists \beta \ [ \ \relpred{}(\alpha,\beta) \wedge 
                        t_{\aautomaton_{\aletter}}(\beta,\proppred{p}(\beta) \ ) 
                        \ ].  
   \]
\end{example}

\begin{table}
\begin{center}
\begin{picture}(100,80)

\put(15,80){ $ \aautomaton_{\aletter}  $ }

\node[Nmarks=i](q0)(20,70){$\astate_0$}
\node(q1)(20,40){$\astate_1$}
\node[Nmarks=r](qf)(20,10){$\astate_f$}

\drawedge(q0,q1){$ \overline{\aletter} $}
\drawloop[loopangle=180](q1){$ \aletter, \overline{\aletter} $}
\drawedge(q1,qf){$ \aletter $}
\drawedge[curvedepth=10](q0,qf){$ \aletter $}

\put(75, 80){$ t_{\aautomaton_{\aletter}}( \alpha, \aformulater(\alpha)) $}

\put(60,71){$ {\bf q}_{0, \aformulater}(\alpha) $}

\put(60,59){$
\forall \alpha \beta \
   [ \ \relpred{}(\alpha,\beta) \rightarrow
      {\bf q}_{0, \aformulater}(\alpha) \rightarrow
      {\bf q}_{f, \aformulater}( \beta) \ ] $ }

\put(60,53){$
\forall \alpha \beta \
   [ \ \relpred{}(\beta,\alpha) \rightarrow
      {\bf q}_{0, \aformulater}(\alpha) \rightarrow
      {\bf q}_{1, \aformulater}(\beta) \ ] $ }

\put(60,41){$
\forall \alpha \beta \
   [ \ \relpred{}(\alpha,\beta) \rightarrow
    {\bf q}_{1, \aformulater}(\alpha) \rightarrow
    {\bf q}_{1, \aformulater}(\beta) \ ] 
$
}

\put(60,35){$
\forall \alpha \beta \
   [ \ \relpred{}(\beta,\alpha) \rightarrow
    {\bf q}_{1, \aformulater}(\alpha) \rightarrow
    {\bf q}_{1, \aformulater}(\beta) \ ] 
$
}

\put(60,23){$
\forall \alpha \beta \
   [ \ \relpred{}(\alpha,\beta) \rightarrow
      {\bf q}_{1, \aformulater}(\alpha) \rightarrow
      {\bf q}_{f, \aformulater}(\beta) \ ]
$
}

\put(60,11){$
\forall \alpha \ 
   [ \ {\bf q}_{f, \aformulater}(\alpha) \rightarrow \aformulater(\alpha) \ ]
$
}

\end{picture}
\end{center}
\caption{The K5 automaton and $t_{\aautomaton_{\aletter}} 
                                (\alpha, \aformulater(\alpha))$
         for arbitrary $ \aformulater(\alpha) $ } 
\label{table-k5}
\end{table}

Since we perform the introduction of new symbols subformula-wise,
it is possible to put the translation of the automaton
outside of the translation of the modal formula. 
(We will use this in Section~\ref{section-ku}).
At the position where 
$ t_{\aautomaton_{\aletter}}( \alpha, t(\aformulabis,\alpha,\beta)) $ is
translated, only $ {\bf q_{0, \aformulabis}}(\alpha) $ needs to be inserted
where $q_0$ is the initial state of $\aautomaton_{\aletter}$.
The rest of the (translation of the) automaton can be put elsewhere.

\paragraph{Extension with nominals.}
The map $T_{\asystem}$ can be obviously
extended to admit nominals in the language of the regular grammar logics with converse. 
The treatment of nominals can be done
in the usual way by extending the definition of $t$ as follows:
$
t(\anominal, \alpha, \beta) \egdef \nompred{\anominal} = \alpha$
where $\nompred{\anominal}$ is a constant associated with the nominal $\anominal$.
The target first-order fragment is $\GF^2$ with constants and identity.
For instance, nominal tense logics with transitive frames 
(see e.g.,~\cite{Areces&Blackburn&Marx00}), 
and description logics with transitive roles and converse
(see e.g.,~\cite{Sattler96}), can be translated into $\GF^2[=]$ with constants
in such a way. Additionnally, by using~\cite[Sect. 4]{Blackburn&Marx02}
regular grammar logics with converse augmented with Gregory's ``actually'' 
operator~\cite{Gregory01} can be translated into such nominal tense logics.

\paragraph{Relationships with first-order logic over finite words.}
The method of translating finite automata into first-order formulas by
introducing unary predicate symbols for the states, is reminiscent to the
characterization of regular languages in terms of Monadic Second-Order Logic
over finite words, namely SOM[+1], see e.g.~\cite{Straubing94}. Similarly, the
class of languages with a {\em finite} syntactic monoid is precisely 
the class of regular
languages. Our encoding into $\GF^2$ is quite specific since
\begin{itemize}
\itemsep 0 cm
\item we translate into an \exptime \ fragment of FOL, namely $\GF^2$, neither
  into full FOL nor into a logic over finite words;
\item we do not encode regular languages into $\GF^2$ but rather modal logics
  whose frame conditions satisfy some regularity conditions, expressible
  in $\GF$ with built-in relations~\cite{Ganzinger&Meyer&Veanes99};
\item not every regularity condition can be encoded by our method since we
  require a closure condition.
\end{itemize}
Hence, the similarity between the encoding of regular languages into SOM[+1]
and our translation is quite superficial. The following argument provides
some more evidence 
that the similarity exists only at the syntactic level. 
The class of regular languages definable with the first-order theory of
SOM[+1] is known as the class of star-free languages (their syntactic monoids
are finite and aperiodic), see e.g.~\cite{Perrin90}. However,
the regular language $\alang_{\asystem}(\aletter)
= (\aletterbis \cdot \aletterbis)^* (\aletter \cup \epsilon)$ obtained
with the regular semi-Thue
system $\asystem = \set{\aletter \rightarrow \aletterbis \aletterbis \aletter, \aletter
  \rightarrow \epsilon}$ produces a regular grammar logic with converse that
can be translated into $\GF^2$ by our method.  Observe that
the language $(\aletterbis \cdot
\aletterbis)^* (\aletter \cup \epsilon)$
is not star-free, see e.g.~\cite{Pin94}. By contrast, $(\aletter \cdot
\aletterbis)^*$ is star-free but it is not difficult to show that there is no
context-free semi-Thue system $\asystem$ such that 
$\alang_{\asystem}(\aletter)
= (\aletter \cdot
\aletterbis)^*$ since $\aletter$ is not in $ (\aletter \cdot
\aletterbis)^*$. As a conclusion, our translation into $\GF^2$ is based
on principles different from those between star-free regular languages
and first-order logic on finite words. Other problems on (tree) automata translatable into
classical logic can be found in~\cite{Verma03}.

\subsection{Satisfiability Preservation}
\label{correctness-of-translation}

We show that map $ T_{\asystem} $ preserves satisfiability. 
First, we introduce some notation. 
A {\em first-order model} is denoted by $\pair{W}{V}$ where $W$ is a non-empty
set and $V$ maps unary [resp. binary] predicate symbols into subsets of 
$W$ [resp. $W \times W$].
Given a first-order model $ \pair{W}{V}, $ we define
$ V(\aletter)$ for every $\aletter \in \aalphabet$ as follows: \\
$V(\aletter) = \left\{\begin{array}{l}
          V(\relpred{\aletter}) \ \mbox{if $\aletter \in \aalphabet^+$} \\
          V(\relpred{\aletter})^{-1} \ \mbox{if $\aletter \in \aalphabet^-$} 
          \end{array}
         \right.$\\[0.2em] 
%
%
The following, rather technical, lemma states roughly the following:
Suppose we have a first-order model
$ \pair{W}{V} $ containing some point $ w \in W, $ 
such that in every point $ v, $ reachable from $ w $ through a path that
is accepted by the automaton $ \aautomaton $  
the formula $ \aformulater( \alpha ) $ is true, 
then we can extend $ V $ in such a way, that
the new model $ \pair{W}{V'} $ will satisfy the translation
$ t_{\aautomaton}(\alpha, \aformulater) $ in $ w. $ 

\begin{lemma}
   \label{essential-property-of-automaton-translation-2}
   Let $ \aautomaton $ be an NDFA and
   $ \aformulater(\alpha) $ be a first-order formula with
   one free variable $ \alpha. $ 
   Let $ \amodel = \pair{W}{V} $ be a first-order structure 
   not interpreting any of the fresh symbols introduced by
   $ t_{\aautomaton}(\alpha,\aformulater)$ (those of the form ${\bf
   q}_{\aformulater}$ for each state $q$ of $\aautomaton$).
   Then there is an extension $ \amodel' = \pair{W}{V'} $ of 
   $ \amodel, $ such that for every $ w \in W, $ satisfying (I) below,
   $ w $ also satisfies
   \[ \amodel', \ v[\alpha \leftarrow w] \models 
          t_{\aautomaton}(\alpha, \aformulater ). \]
   \begin{description}
   \itemsep 0 cm
   \item[(I)]   
   For every word 
   $ \aletterbis_1 \cdots \aletterbis_n \in \aalphabet^{*} $ that is accepted 
   by $ \aautomaton, $ for
   every sequence $ w_1, \ldots, w_n $ of elements of $W$ such that
   \[  \pair{w}{w_1} \in V(\aletterbis_1), \ \
       \pair{w_1}{w_2} \in V(\aletterbis_2), \ \ldots, \ \
       \pair{w_{n-1}}{w_n} \in V(\aletterbis_n), \]
       we have
   $ \amodel, v[ \alpha \leftarrow w_n ] \models \aformulater(\alpha). $
   \end{description}
\end{lemma}

\begin{proof}
   We want to extend $ V $ in such a way that the added interpretations
   for the symbols $ {\bf q}_{\aformulater} $ simulate $ \aautomaton. $ 
   This can be obtained by the following extension:
   For all $ w \in W$ and  all states $ \astate $  of $ \aautomaton, $ 
   we set $ w \in V'( {\bf \astate}_{\aformulater} ) $ iff
   for every  word $\aletterbis_1 \cdots \aletterbis_n  \in \aalphabet^{*} $ 
   such that 
   there is an accepting state $ q_f $ of $ \aautomaton, $ such that 
   $ q_f \in \delta^{*}( \astate, \aletterbis_1 \cdots \aletterbis_n)$, 
   and for every sequence $ w_1, \ldots, w_n $ of
   elements of $ W, $ s.t.
   \[  \pair{w}{w_1} \in V(\aletterbis_1), \ \
       \pair{w_1}{w_2} \in V(\aletterbis_2), \ \ldots, \ \
       \pair{w_{n-1}}{w_n} \in V(\aletterbis_n), \]
   we have 
   \[ \amodel, v [ \alpha \leftarrow w_n ] \models
      \aformulater( \alpha ). \]
   Here $ \delta^{*} $ is the natural extension of 
   $ \delta $ to words
   over $ \Sigma^{*}. $ 

   Write $ \aautomaton = \triple{Q}{s, F}{\delta}$. 
   It is easy to check (but tedious to write out because of 
   the size of the statements involved) that for every $w \in W$ satisfying
   (I) in the statement of the lemma, we have
   \begin{itemize}
   \itemsep 0 cm
   \item
      for the initial state $ s, $ 
      \[ \amodel', v [ \alpha \leftarrow w ] \models {\bf s}_{\aformulater}(\alpha). \]   
   \item
      for each $ q \in Q, $ for each $ \aletter \in \aalphabet, $ 
      for each $ r \in \delta( q, \aletter), $ 
      \[ \amodel' \models
         \forall \alpha \beta \ [ 
         \ t_{\aletter}(\alpha, \beta) \rightarrow
                       {\bf q}_{\aformulater}(\alpha) \rightarrow 
                       {\bf r}_{\aformulater}(\beta) \ ]. \] 
   \item
      for each $q,r  \in Q$, such that
      $ r \in \delta(q, \epsilon), $ 
      \[ \amodel' \models
         \forall \alpha \ [ 
         \ {\bf q}_{\aformulater}(\alpha) \rightarrow {\bf r}_{\aformulater}(\alpha) \ ]. \]
   \item 
      for each final state $ q \in F$, 
      \[ \amodel' \models
         \forall \alpha \ 
         [ \ {\bf q}_{\aformulater}(\alpha) \rightarrow \aformulater(\alpha) \ ]. \]
   \end{itemize}   
   $ \amodel' $ and $ \amodel $ agree on all formulas that
   do not contain any symbols introduced by 
   $ t_{\aautomaton}( \alpha, \aformulater )$. 
\end{proof}

\noindent
Next follows the main theorem about satisfiability preservation.

\begin{theorem} 
   \label{theorem-main-translation}
   Let $ \aalphabet $ be an alphabet with converse mapping
   $ \overline{\cdot}, $ let 
   $ \asystem $ be a regular semi-Thue system closed under converse
   over $\aalphabet$, and let 
   $\aformula \in {\cal L}^{\aalphabet}$ be a modal formula. Then, 
   \begin{description}
   \itemsep 0 cm
   \item[(I)]
      $\aformula$ is $ S $-satisfiable iff 
   \item[(II)]
      $T_{\asystem}(\aformula)$ is satisfiable in FOL.
   \end{description} 
\end{theorem}

\noindent
The proof relies on the 
regularity of the languages $ \alang_{\asystem}(\aletter) $
and on Theorem~\ref{theorem-closure-is-lazy}.

\begin{proof}
We first prove {\bf (I) $\rightarrow$ (II)}. 
Assume that $\aformula$ is $ \asystem $-satisfiable. 
This means that there exists a 
$ \pair{\aalphabet}{\overline{\cdot} \ } $-model
$\amodel = \triple{W}{R}{V} $
with a $ w \in W$ such that $\amodel, w \models \aformula$ 
and $ \pair{W}{R} $ satisfies $ \asystem. $ 
We need to construct a model $ \amodel' $ (noted $\amodel_n$ in the sequel) of $ T_{\asystem}( \aformula ). $ 
In order to do this, we first construct an incomplete 
interpretation $\amodel_0 = \pair{W}{V_0}, $ and after that
we will complete it through successive applications of 
Lemma~\ref{essential-property-of-automaton-translation-2}.

\begin{itemize}
\itemsep 0 cm
\item for every $\aletter \in \aalphabet^{+}$, \ 
   $V_0(\relpred{\aletter}) \egdef R_{\aletter}$;
\item for every propositional variable $ \avarprop $, we set
   $ V_0(\proppred{\avarprop}) \egdef V(\avarprop)$.
\end{itemize}
We now have a model interpreting 
the symbols introduced by $ t( \aformulabis, \alpha, \beta ), $ but
not the symbols introduced by $ t_{\aautomaton}( \alpha, \aformulabis ). $ 
In order to complete the model construction, 
we order the box-subformulas of $ \aformula $ in a sequence
$ \bo{\aletter_1} \aformulabis_1, \ldots, \bo{\aletter_n} \aformulabis_n $ such 
that every box-subformula is preceeded by all its box-subformulae.
Hence, $i < j$ implies that $\bo{\aletter_j} \aformulabis_j$ is not a
subformula
of $\bo{\aletter_i} \aformulabis_i$.
Then we iterate the following construction ($1 \leq i \leq n$): 
\begin{itemize}
\item
   $ \amodel_{i} = \pair{W}{V_{i}} $ is obtained from
   $ \amodel_{i-1} = \pair{W}{V_{i-1}} $ by applying the construction
   of Lemma~\ref{essential-property-of-automaton-translation-2} 
   on $ \aautomaton_{\aletter_i} $ and $ t( \aformulabis_i, \alpha, \beta). $ 
\end{itemize} 
Then $ \amodel_n = \pair{W}{V_n} $ is our final model. 
Roughly speaking, $V_{i}$ is equal to $V_{i-1}$ extended to the unary
predicate symbols of the form
${\bf q}_{\aformulabis_i}$ with $q$ a state of $\aautomaton_{\aletter_i}$.
The values of the other predicate symbols remain constant, that is:
\begin{itemize}
\itemsep 0 cm
\item for every $\aletter \in \aalphabet^+$, $V_0(\relpred{\aletter}) = \cdots =
                                            V_n(\relpred{\aletter})$;
\item for every propositional variable $\avarprop$ in $\aformula$, 
      $V_0(\proppred{\avarprop}) = \cdots =
                                            V_n(\proppred{\avarprop})$.
\end{itemize}
Additionally, for every $j \in \set{1, \ldots, n}$, for every state $q$ of
$\aautomaton_{\aletter_j}$, $V_{j}({\bf q}_{\aformulabis_j}) = 
V_{j+1}({\bf q}_{\aformulabis_j}) = \ldots
= V_{n}({\bf q}_{\aformulabis_j})$.

We show by induction that for every subformula $\aformulabis $
of $ \aformula, $ for every $x \in W $, 
for every valuation $ v, $ \ \ 
   $ \amodel, x \models \aformulabis $ implies 
   $ \amodel_n, v[\alpha \leftarrow x] \models 
      t(\aformulabis, \alpha, \beta). $
Here $v [\alpha \leftarrow x]$ denotes the valuation $v'$ obtained from $v$ by 
putting
$v'(\beta) \egdef v(\beta)$ and $v'(\alpha) = x$.
We treat only the modal cases, because the propositional cases
are trivial. 

\begin{itemize} 
\itemsep 0 cm
\item
   If $ \aformulabis $ has form 
      $ \bo{\aletter} \aformulabis' $ with $ \aletter \in
   \aalphabet, $ then 
   $ t( \bo{\aletter} \aformulabis', \alpha, \beta) = 
        t_{\aautomaton_{\aletter}}( \alpha, t(\aformulabis', \alpha,\beta)). $ 

   We have that for every
   %
   word $\aletterbis_1 \cdots \aletterbis_l$ accepted by $\aautomaton_{\aletter}$, 
   for every sequence $ w_1, \ldots, w_l \in W_{n} $ s.t.
   \[ \pair{x}{w_1} \in V_n(\aletterbis_1), \ \ 
      \pair{w_1}{w_2} \in V_n(\aletterbis_2), \ \ldots \ , 
      \pair{w_{l-1}}{w_l} \in V_n(\aletterbis_l),  \]
   also 
   \[
    \pair{x}{w_1} \in R_{\aletterbis_1}, \
    \pair{w_1}{w_2} \in R_{\aletterbis_2}, \ \cdots \ ,
    \pair{w_{l-1}}{w_l} \in R_{\aletterbis_l},      
   \]
   by construction of $V_0, V_1, \cdots, V_n$.  
   Because $\amodel$ satisfies $\asystem$,  
   $\pair{x}{w_l} \in R_{\aletter}$
   (by 
   Theorem~\ref{theorem-three-characterizations}((I) $\rightarrow$ (III))),
   which again implies $\pair{x}{w_l} \in V_n(\aletter)$, 
   by construction of the $V_i$. 
   Therefore, we have $ \amodel, w_l \models \aformulabis'$. 
   By the induction hypothesis, we have
   $ \amodel_n, v [ \beta \leftarrow w_l ] \models 
         t( \aformulabis',\beta,\alpha)$. 
   Let $ n' $ be the position of $ \aformulabis' $ in the enumeration 
   of box-subformulae 
   $ \bo{\aletter_1} \aformulabis_1, \ldots, \bo{\aletter_n} \aformulabis_n. $ 
   It is easily checked that 
   \[ \amodel_{n'}, v [ \beta \leftarrow w_l ] \models
      t( \aformulabis', \beta, \alpha). \]
   Now we have all ingredients of 
   Lemma~\ref{essential-property-of-automaton-translation-2} 
   complete, and it follows that
   \[ {\amodel}_{n'}, v [ \alpha \leftarrow x ] \models
         t_{\aautomaton_{\aletter}}(\alpha, t(\aformulabis', \alpha, \beta)). \]

Since $\amodel_n$ is a conservative extension $\amodel_{n'}$, we also get

   \[ {\amodel}_{n}, v [ \alpha \leftarrow x ] \models
         t_{\aautomaton_{\aletter}}(\alpha, t(\aformulabis', \alpha, \beta)). \]

\item
   If $\aformulabis$ has form $ \diam{\aletter} \aformulabis'$, then 
    there is a $y$ such that $\pair{x}{y} \in R_{\aletter}$ 
   and 
   $\amodel, y \models \aformulabis' $. By definition of $ V_0$, 
   we have $ \pair{x}{y} \in V_0(\aletter)$ 
   and therefore $ \pair{x}{y} \in V_n(\aletter). $ 
   By the induction hypothesis, ${\amodel}_n, v [ \beta \leftarrow y ] \models
         t( \aformulabis', \beta, \alpha)$.
   Hence, 
   \[ \amodel_n, v[\alpha \leftarrow x] \models 
      \exists \beta \ [ \ t_{\aletter}(\alpha,\beta) \wedge
      t(\aformulabis', \beta,\alpha) ]. \]  
\end{itemize}

\noindent
{\bf (II) $\rightarrow$ (I)}
Suppose that $T_{\asystem}(\aformula)$ is FOL-satisfiable.
This means that there exist a FOL model $\amodel = \pair{W}{V}$ and a 
valuation $v$ such that $\amodel, v \models T_{\asystem}(\aformula)$.
We construct a model $ \amodel' $ of $ \aformula $ in two stages:
First we construct $ \amodel'' = \triple{W''}{R''}{V''} $ as follows:
\begin{itemize}
\itemsep 0 cm
\item $ W'' = W$. 

\item For every $\aletter \in \aalphabet$, 
      $R''_{\aletter} = V(\aletter)$.
\item 
   For every propositional variable $\avarprop$, $V''(\avarprop) = V({\bf p})$. 
\end{itemize}
Then define $ \amodel' = \triple{W'}{R'}{V'} $ where $ R' $ is defined
from $ \pair{W'}{R'} = C_S(\pair{W''}{R''} ) $ and $ V' = V''$.
Here $ C_S $ is the closure operator, defined in 
Definition~\ref{definition-closure-operator}.
Intuitively, we construct $ \amodel' $ by copying $ W $ and the
interpretation of the accessibility relations from $ \amodel, $ 
and applying $ C_{\asystem} $ on it. The constructions imply
that $ W' = W. $ 
By definition of $ C_{\asystem}, \ \ \  \amodel' $ is an ${\asystem}$-model,
and by Lemma~\ref{lemma-closure-converse}, $ \pair{W''}{R''} $ is a 
$\pair{\aalphabet}{\overline{\cdot}}$-frame. 
We now show by induction that  
for every subformula $ \aformulabis $ of $ \aformula, $ \ 
$\amodel, v \models t(\aformulabis, \alpha, \beta)$
implies $\amodel', v(\alpha) \models \aformulabis$.

\begin{itemize}
\itemsep 0 cm
\item
   If $ \aformulabis $ has form $\diam{\aletter} \aformulabis'$,
   then $\amodel, v \models t(\diam{\aletter} \aformulabis', \alpha, \beta)$, 
   that is
   $\amodel, v \models \exists \beta \ [ \ t_{\aletter}(\alpha, \beta)
                                         \wedge t(\aformulabis',
                                         \beta, \alpha) \ ]. $ 

   This means there is a $ y \in W, $ such that
   $\amodel, v [ \beta \leftarrow y ] \models t_{\aletter}(\alpha,\beta), $ 
   and    
   \[\amodel, v [ \beta \leftarrow y ] \models 
      t(\aformulabis', \beta, \alpha). \] 
   By the induction hypothesis, 
   $ \amodel', y \models \aformulabis'$. 
   It follows from the definition of $R'$, 
   using the fact that $ C_S $ is increasing (by its definition),
   that $\pair{x}{y} \in R'_{\aletter}$,
   so we have $\amodel', x \models \diam{\aletter} \aformulabis. $ 
\item
   If $\aformulabis$ has form 
   $\bo{\aletter} \aformulabis'$, then suppose
   $ \amodel, v \models t_{\aautomaton_{\aletter}} 
        ( \alpha, t( \aformulabis', \alpha, \beta)  ) $.
   One can establish that  
  for every word $\aletterbis_1 \cdots \aletterbis_l$ 
   accepted by $\aautomaton_{\aletter}$, for every sequence 
   $w_1, \ldots, w_l$ of elements of $W$, for which it is the
   case that 
   \[ \pair{v(\alpha)}{w_1} \in V(\aletterbis_1), \ 
      \pair{w_1}{w_2} \in V(\aletterbis_2), \ \ldots, 
      \pair{w_{l-1}}{w_l} \in V(\aletterbis_l), \]
   the following holds
   \[ \amodel, v [ \alpha \leftarrow w_l ] \models 
      t( \aformulabis', \alpha, \beta). \]
   Indeed,
   $ \amodel, v  \models {\bf s}(\alpha), $ 
   for the initial state $ s $ of $ \aautomaton_{\aletter}$. 
   It is easy to show by induction that the following holds:
   Let $\aletterbis_1 \cdots \aletterbis_l$ be some word over $\aalphabet^{*}$. 
   Let $ \delta^{*} $ be the natural extension of $ \delta $ to words. 
   Let $ q $ be a state of $\aautomaton_{\aletter}$ such that 
   $ q \in \delta^{*}( s, \aletterbis_1 \cdots \aletterbis_l), $ for the initial
   state $ s \in Q$. 
   Then for every sequence $ w_1, \ldots, w_l $ of elements 
   of $W$ such that 
   \[ \pair{v(\alpha)}{w_1} \in V(\aletterbis_1), \ \ 
      \pair{w_1}{w_2} \in V(\aletterbis_2), \ \ \ldots, \ \ 
      \pair{w_{l-1}}{w_l} \in V(\aletterbis_l), \]
   it must be the case that 
   $ \amodel, v[ \alpha \leftarrow w_l ] \models {\bf q}(\alpha). $  
   Then the result follows from the fact that 
   $ \amodel, v [ \alpha \leftarrow w_l ] \models
     {\bf q}(\alpha) \rightarrow \aformulater(\alpha), $ 
   for each accepting state $q$ of $\aautomaton_{\aletter}$. \\

   Now assume that in $ \amodel', $ we have a world
   $ y $ for which $ R'_{\aletter}(x,y)$. 
   Then, using Theorem~\ref{theorem-closure-is-lazy},
   there is a word $ w $ that is accepted by
   $ \aautomaton_{\aletter}, $ such that 
   $ R''_w(x,y). $ 
   By the above property, we have
   $ \amodel, v [ \alpha \leftarrow y ] \models 
     t( \aformulabis', \alpha, \beta). $ 
   By the induction hypothesis, we have
   $ \amodel', y \models \aformulabis'. $ 
\end{itemize} 
\end{proof}

\noindent
The uniformity of the translation allows us to establish forthcoming 
Theorem~\ref{theorem-main-result}.
We first define the general satisfiability problem for regular grammar logic 
with converse,
denoted by GSP(REG$^{c}$), as follows:
\begin{description}
\itemsep 0 cm
\item[input:] A semi-Thue system $ \asystem $ with converse, in 
   which either all rewrite rules are left-linear, or
   all rewrite rules are right-linear, and 
   an ${\cal L}^{\aalphabet}$-formula $\aformula$;
\item[question:] is $\aformula$ $S$-satisfiable?
\end{description}

We need to restrict the form of the semi-Thue system to a 
form from which the automata $ \aautomaton_{\aletter} $ can
be computed. 
Even if one knows that some language $ \alanguage $ is regular,
then there is no effective way of obtaining an NDFA 
for $ \alanguage. $ This is a consequence of Theorem~2.12 (iii) in
\cite{Rozenberg&Salomaa1994}. 

\begin{theorem} \label{theorem-main-result} \ 
\begin{description}
\item[(I)] The $\asystem$-satisfiability problem is in \exptime \ 
           for every regular  semi-Thue system with converse.
\item[(II)] GSP(REG$^{c}$) is \exptime-complete.
\end{description}
\end{theorem}

Theorem~\ref{theorem-main-result}(II) is  stronger than 
Theorem~\ref{theorem-main-result}(I) because 
GSP( REG$^{c}$) covers the satisfiability problems
for all regular grammar logics with converse. 
Theorem~\ref{theorem-main-result}(I) is a corollary of
Theorem~\ref{theorem-main-translation}. 
The lower bound in Theorem~\ref{theorem-main-result}(II) is easily
obtained by observing that there exist known regular grammar logics (even without converse)
that are already \exptime-complete, e.g. K with the universal modality. 
The upper bound in Theorem~\ref{theorem-main-result}(II) is a
consequence of the facts that
$T_{\asystem}(\aformula)$ can be computed in logarithmic space 
in $\length{\aformula} +
\length{\asystem}$ 
and the guarded fragment has 
an \exptime-complete 
satisfiability problem when the arity of the predicate symbols 
is bounded by some fixed $k \geq 2$ \cite{Gradel99c}.
We use here the fact that one needs only logarithmic space 
to build  a finite automaton
recognizing the language of  a right-linear [resp. left-linear] grammar.

\paragraph{Extensions to context-free grammar logics with converse.}
When $\asystem$ is a context-free semi-Thue system with converse, 
$\asystem$-satisfiability can be encoded as for the case of 
regular semi-Thue systems with converse by adding an argument 
to the predicate symbols of the form ${\bf q}_{\aformulabis}$.
The details are omitted here but we provide the basic intuition.
Each language $\alang_{\asystem}(\aletter)$ is context-free
and therefore there is a pushdown automaton (PDA) $\aautomaton$ recognizing it. 
The extra argument for the ${\bf q}_{\aformulabis}$s represents
the content of the stack and the map $ t_{\aautomaton}( \alpha, \aformulater
)$
can be easily extended in the presence of stacks. 
For instance, the stack content $\aletter \aletter \aletterbis$ can be
represented by the first-order term
$\aletter(\aletter(\aletterbis(\epsilon)))$
with the adequate arity for the function symbols $\aletter$, $\aletterbis$, 
and $\epsilon$.
Suppose we have the following transition rule: if the PDA is in state $q$,
the current input symbol is $\aletter$, and the top symbol of the stack 
is $\aletterbis_0$, then the new state is $q'$ and $\aletterbis_0$ is replaced
by $\aletterbis_1 \cdots \aletterbis_n$ on the top of the stack. 
This rule is encoded in FOL as follows:
$$
\forall \ \alpha, \beta, \gamma, \
(t_{\aletter}(\alpha,\beta)
 \Rightarrow
 (
 {\bf q}(\alpha, \aletterbis_0(\gamma))
 \Rightarrow
 {\bf q'}(\beta, \aletterbis_1(\ldots \aletterbis_n(\gamma) \ldots ))
 )).
$$
The translation $T_{\asystem}$ is then defined with the context-free version
of $t_{\aautomaton}( \alpha, \aformulater)$. Satisfiability preservation is also guaranteed but the first-order
fragment
in which the translation is performed (beyond $\GF$) is not anymore decidable. Hence,
although this provides a new translation of context-free grammar logics
with converse, from the point of view of effectivity, this is not better than
the relational translation which is also known to be possible when  
$\asystem$ is a context-free semi-Thue system with converse.

\section{Alternative Proofs of the \exptime \ Upper Bound}
\label{section-alternative}

In this section, we provide two alternative ways to show that GSP(REG$^{c}$)
is in \exptime. We believe that not only this sheds some new light to the proof of
Section~\ref{section-2gf-translation} but also it emphasizes the peculiarities
of the class of regular grammar logics with converse. Observe that filtration-like techniques
might also establish decidability of  logics from GSP(REG$^{c}$), if not to the whole fragment.
However, with such a technique the size of the built models is usually at
least exponential in the size of the formulae, so we might
get at best an \nexptime \ upper bound. That is why we did not develope
further here this kind of proof.

\subsection{Converse PDL with Automata}
\label{section-acpdl}
In~\cite{Demri01}, it is shown how to translate the general satisfiability
problem for regular grammar logics without converse into satisfiability for PDL but
this map was not logarithmic space because given a regular grammar, equivalent
regular expressions can be of exponential size (see e.g.~\cite{HopcroftUllman1979}).
That is why PDL with automata (APDL), see e.g.~\cite{Harel&Kozen&Tiuryn00},  
has been considered in~\cite{Demri01} in order to 
obtain a logarithmic space transformation into an \exptime \ logic. 
Similarly, it is  possible to define a logarithmic space transformation from 
GSP(REG$^{c}$) into $\GF^2$ by first 
translating  GSP(REG$^{c}$) into ACPDL (converse PDL with automata) and then by translating a fragment of ACPDL into
$\GF^2$. ACPDL is an extension of APDL where the set of atomic programs is a countably
infinite alphabet with converse mapping of the form 
$\set{\aletter_i: i \in \Nat^*} \cup \set{\overline{\aletter_i}: i \in \Nat^*}$
In the ACPDL models, we have $R_{\overline{\aletter}} = R_{\aletter}^{-1}$. 
The translation from  GSP(REG$^{c}$) into ACPDL is mainly based on the step
translating $\bo{\aletter} \aformulabis$ into 
         $\bo{\aautomaton_{\aletter}} t(\aformulabis)$ where $\aautomaton_{\aletter}$ is a ACPDL-automaton
       accepting the regular language $\alang_{\asystem}(\aletter)$. When $\asystem$ is either right-linear or
       left-linear, $\aautomaton_{\aletter}$ can be computed in polynomial-time in $\length{\asystem}$.
Observe that ACPDL is more expressive than the class of regular grammar logics with converse
since there is no
context-free semi-Thue system $\asystem$ such that 
$\alang_{\asystem}(\aletter_1)
= (\aletter_1 \cdot \aletter_2)^*$.

By mimicking the translation into $\GF^2$ from Section~\ref{section-transformation}, one cannot
translate full ACPDL into $\GF^2$. Indeed, a syntactic restriction on ACPDL formulae
 similar to the existence of
a closure operator on $\aalphabet$-frames can be defined so that such a ACPDL fragment can be
translated into $\GF^2$. Such a fragment contains the translated formulae from GSP(REG$^c$)
However, we omit here the definition of the translation since
then the translation is not so much more informative that the one from Section~\ref{section-transformation}.

The ACPDL fragment in question contains the formulae with ACPDL automata
$\aautomaton_1, \overline{\aautomaton_1}, \ldots, \aautomaton_n, \overline{\aautomaton_n}$ 
built over the atomic
programs $\set{\aletter_1, \ldots, \aletter_n} \cup \set{\overline{\aletter_1},  \ldots, \overline{\aletter_n}}$
satisfying the properties below:
\begin{itemize}
\itemsep 0 cm
\item the words of $\alang(\aautomaton_i)$ are the reverse words of
$\alang(\overline{\aautomaton_i})$. 
\item for every $i \in \set{1, \ldots,n}$, $\aletter_i \in \alang(\aautomaton_i)$.
\item for every word $\astring \in \alang(\aautomaton_i)$ [resp. $\astring \in \alang(\overline{\aautomaton_i})$],
            \begin{enumerate}
            \itemsep 0 cm
            \item for every occurrence $\aletter_j$, 
                  the word obtained from $\astring$ by replacing that occurrence of $\aletter_j$ 
                  by any word   in  $\alang(\aautomaton_j)$ is also in $\alang(\aautomaton_i)$ [resp. 
                  $\alang(\overline{\aautomaton_i})$];
            \item for every occurrence $\overline{\aletter_j}$, 
                  the word obtained from $\astring$ by replacing that occurrence of 
                  $\overline{\aletter_j}$ 
                  by any word   in  $\alang(\overline{\aautomaton_j})$ is also in 
                  $\alang(\aautomaton_i)$
                   [resp. $\alang(\overline{\aautomaton_i})$].
            \end{enumerate}
\end{itemize}

%
%
%

\subsection{Multimodal K$_t$ + $\bo{U}$}
\label{section-ku}

We have defined in Section~\ref{section-transformation} an almost structure-preserving map from
regular grammar logics with converse into $\GF^2$.  Below, 
we provide hints to understand how this map can be turned into a map into
the multimodal logic K$_t$ with forward and backward modalities
$\set{\bo{i}, \bo{i}^{-1}, \diam{i}, \diam{i}^{-1} : i \geq 1}$\footnote{
$\diam{i}^{-1}$ is also noted $P_i$ (existential past-time operator)
and $\bo{i}^{-1}$ is also noted $H_i$ (universal past-time operator).}
 augmented with the universal 
modality $\bo{U}$. 
More interestingly, this logic can be then viewed as an \exptime-complete 
 pivot logic between 
regular grammar logics with converse and $\GF^2$ and 
many decision procedures exist for it (see e.g.,
\cite{Donini&Massacci00,Hustadt&Schmidt00,Baader&Tobies01}).

Let $\asystem$ be a regular semi-Thue system with converse and
let $\aformula$ be an ${\cal L}^{\Sigma}$-formula in negation normal form (NNF).
Let us define  $t(\aformula)$ by induction on
the subformulae of $\aformula$. 

\begin{itemize}
\itemsep 0 cm
\item $t(l) \egdef l$ for every literal $l$; 
\item $t(\aformulabis \wedge \aformulabis') \egdef
       t(\aformulabis) \wedge t(\aformulabis')$ 
(similar for $\vee$);
\item $t(\diam{\aletter} \aformulabis) \egdef 
      \diam{i_{\aletter}} t(\aformulabis)$ where
      $i_{\aletter}$ is a modal index associated  
      with $\aletter \in \aalphabet^{+}$;
\item $t(\diam{\overline{\aletter}} \aformulabis ) \egdef 
      \diam{i_{\aletter}}^{-1} t(\aformulabis )$ 
      with $\aletter \in \aalphabet^{+}$;
\item $t(\bo{\aletter} \aformulabis ) \egdef
       \avarprop_{s, \aformulabis}$
      where $\avarprop_{s, \aformulabis}$ is a propositional variable associated
      with the initial state $s$ of $\aautomaton_{\aletter}$, 
      $\aletter \in \aalphabet$,  and $\bo{\aletter} \aformulabis \in \subf{\aformula}$.
\end{itemize}

More generally, for every $\bo{\aletter} \aformulabis \in \subf{\aformula}$ and for
every
$\astate \in \aautomaton_{\aletter}$, we shall introduce a new propositional 
variable $\avarprop_{\astate, \aformulabis}$.

As done in Section~\ref{section-2gf-translation}, 
for every $\aletter \in \aalphabet$, for every $\aformulabis$ such that
$\bo{\aletter} \aformulabis$ occurs in $\aformula$, we construct
formulas $t(\aautomaton_{\aletter}, \aformulabis )$. 
The conjunctions contain the following formulas: 

\begin{itemize}
\itemsep 0 cm 
\item
   For each $\aletterbis \in \aalphabet^{+}, \ \
      q, r \in Q_{\aletter}, $ 
   if $ r \in \delta_{\aletter}(q, \aletterbis)$, then
   the formula \\ 
  $\bo{U}(\avarprop_{q, \aformulabis} \Rightarrow \bo{i_{\aletterbis}} \avarprop_{r, \aformulabis})$
   is present. 
\item
For each $\aletterbis \in \Sigma^{-}, \ \
      q, r \in Q_{\aletter}, $ 
   if $ r \in \delta_{\aletter}(q, \aletterbis)$, then
   the formula \\
   $\bo{U}(\avarprop_{q, \aformulabis} \Rightarrow \bo{i_{\aletterbis}}^{-1}
   \avarprop_{r, \aformulabis})$
   is present.
\item
   For $ q,r \in Q_{\aletter}$, 
   if $ r \in \delta_{\aletter}(q,\epsilon)$, then
   the formula 
   $\bo{U}(\avarprop_{q, \aformulabis} \Rightarrow  \avarprop_{r, \aformulabis})$
   is present. 
\item
   For each $ q \in F_{\aletter}$, the formula
   $\bo{U}(\avarprop_{q, \aformulabis} \Rightarrow  t(\aformulabis))$
   is present. 
 \end{itemize}

The translation $T_{\asystem}(\aformula)$ is defined as $t(\aformula)
\wedge \bigwedge_{\bo{\aletter} \aformulabis \in \subf{\aformula}} 
t(\aautomaton_{\aletter}, 
\aformulabis )$.
The size of $T_{\asystem}(\aformula)$ is in ${\cal O}(\length{\aformula} + \length{\asystem})$ and
 $T_{\asystem}(\aformula)$ can be computed in logarithmic space
in $\length{\aformula} + \length{\asystem}$.

\begin{theorem} \label{theorem-ku}
Let $\aalphabet$ be an alphabet with converse mapping $ \overline{\cdot}$, 
    $\asystem$ be a regular  semi-Thue system with converse over $\aalphabet$, and
$\aformula \in {\cal L}^{\aalphabet}$. Then,
  $\aformula$ is $\asystem$-satisfiable iff 
  $T_{\asystem}(\aformula)$ is K$_t$ + $\bo{U}$ satisfiable. 
\end{theorem}

The proof is by an easy verification by observing that the first-order formula obtained 
with the relational translation from the
K$_t$ + $\bo{U}$ formula $T_{\asystem}(\aformula)$ is almost
syntactically
equal to the formula $T_{\asystem}(\aformula)$ 
from Section~\ref{section-2gf-translation} when the last three clauses of
Definition~\ref{definition-translation-automaton} are extracted from
  $t_{\aautomaton}( \alpha, \aformulater )$
which is correct when the translation is done subformula-wise.
Theorem~\ref{theorem-main-translation} concludes the proof.
In general, the first-order formulae obtained by relational translation
from K$_t$ + $\bo{U}$ formulae are not in the guarded fragment.
However, the relational translation of a formula $T_{\asystem}(\aformula) $ 
is always in $ \GF^2 $.
Extensions with nominals in ${\cal L}^{\aalphabet}$ and in 
K$_t$ + $\bo{U}$ is obvious  by adding the clause $t(\anominal) = \anominal$ for
each nominal $\anominal$.

%
%

Let us consider CPDL, the version of PDL with $;$ (composition), $\cup$ (nondeterministic choice), $^{*}$  (iteration),
$^{-1}$ (converse).
The set of program constants is $\set{\aconst_1, \aconst_2, \ldots}$.
For additional material on CPDL we refer to \cite{Harel&Kozen&Tiuryn00}.
A formula $\aformula$ of K$_t$ + $\bo{U}$ with modal indices in $\set{1, \ldots,
n}$ can be translated to CPDL by replacing every occurrence of $\bo{i}$ by
$\bo{\aconst_i}$, every occurrence of $\bo{i}^{-1}$ by $\bo{\aconst_i^{-1}}$, and
every occurrence of $\bo{U}$ by $\bo{(\aconst_1 \cup \ldots \cup \aconst_{n+1}
\cup \aconst_1^{-1} \cup \ldots \cup \aconst_{n+1}^{-1})^*}$.   
One can show that $\aformula$ is K$_t$ + $\bo{U}$ satisfiable iff the translated
formula
is CPDL satisfiable. Elimination of the universal modality does not make any problem
because we are dealing with connected models.
The proof is standard (see e.g.~\cite{Fischer&Ladner79,Tuominen90,Goranko&Passy92}).
Hence, by combining this result with Theorem~\ref{theorem-ku}, we obtain a
logarithmic space transformation from GSP(REG$^c$) into CPDL.

\section{Translating Intuitionistic Propositional Logic into $\GF^2$}
\label{section-intuitionistic-logic}

We define a new translation from intuitionistic propositional
logic IPL with connectives $ \rightarrow, \vee, \wedge $ and
$\perp$
(see e.g. details in~\cite{Chagrov&Zakharyaschev97})
into $\GF^2$. 

The translation method we present is technically not difficult since
the translation is obtained by composing G\"odel's translation into S4
with our translation from S4 into the guarded fragment,  
see e.g. \cite{TroelstraSchwichtenberg} for the G\"{o}del translation.
This provides 
another logarithmic space embedding of IPL into a decidable fragment of 
classical logic (see e.g.~\cite{Korn&Kreitz97}.
Our translation could be used as a method for theorem proving
in intuitionistic logic, but it still has to be determined whether
our translation results in an efficient procedure. 
For more direct methods, we refer to the contraction-free calculus
of Dyckhoff \cite{Dyckhoff1992}, or a resolution
calculus, see \cite{Mints88}, which is implemented in
\cite{System:Gandalf}. 

Let $\aformula$ be an intuitionistic formula. We cannot base the 
translation on the 
negation normal form of $ \aformula, $ as we did in 
Section~\ref{section-2gf-translation}, 
because intuitionistic logic does not admit an equivalent 
negation normal form. (For example $ \neg \aformula \vee \aformulabis $ 
is not equivalent to $ \aformula \rightarrow \aformulabis $)
Therefore, we explicitly add the polarity to the translation function.
A similar technique was used in for example \cite{Demri&Gore00b}. 

Before defining the map, we repeat the translation from IPL into S4, 
as given in \cite{TroelstraSchwichtenberg}. 

\begin{definition}
   \label{GoedelS4}
   Function $ t_{S4} $ is defined as follows by recursion on the
   subformulas of $ \aformula. $ 
   \begin{itemize}
   \itemsep 0 cm
   \item
      $ t_{S4}( \bot ) $ equals $ \bot, $ 
   \item
      for a propositional symbol $ \avarprop, $ \ 
      $ t_{S4}(\avarprop) $ equals $ \Box \avarprop, $ 
   \item
      $ t_{S4}( \aformulabis \wedge \aformulabis') $ equals
      $ t_{S4}( \aformulabis ) \wedge t_{S4}( \aformulabis' ), $   
   \item
      $ t_{S4}( \aformulabis \vee \aformulabis' ) $ equals
      $ t_{S4}( \aformulabis ) \vee t_{S4}( \aformulabis' ), $ 
   \item
      $ t_{S4}( \aformulabis \rightarrow \aformulabis' ) $ equals
      $ \Box ( t_{S4}( \aformulabis ) \rightarrow 
                 t_{S4}( \aformulabis' )). $
\end{itemize}
\end{definition}
Translation $ t_{S4} $ takes an intuitionistic formula 
$ \aformula $ and returns an S4-formula, not necessarilly
in NNF.  
Translation $ t_{S4} $ preserves {\em provability}, so 
formula $ \aformula $ is provable in IPL iff $ t_{S4}(\aformula) $ is S4-valid.
A formula $ t_{S4}(\aformula) $ is provable iff its negation
$ \neg t_{S4}( \aformula) $ is unsatisfiable. 
In order to use the methods of Section~\ref{section-2gf-translation}, 
the translation function $ t_{S4} $ has to be modified in such 
a way that it (1) directly constructs the negated formula, and
(2) it constructs a modal formula in negation normal form.
The result is the following modified transformation $ t_{S4}. $ 
It takes an intuitionistic formula $ \aformula $ and the 
polarity $ \pi \in \{ 0, 1 \}, $ which is determined by the context
of the result. 
In order to construct the translation of a formula
$ \aformula, $ one needs to construct $ t_{S4}(\aformula, 0). $ 
Then $\aformula$ is provable in IPL iff 
$ t_{S4}(\aformula,0) $ is unsatisfiable. The new translation
function $ t_{S4} $ is defined by recursion:

\begin{definition}
   \label{GoedelS4modified} \ 
   \begin{itemize}
   \itemsep 0 cm
   \item
      $ t_{S4}( \bot, 1 ) $ equals $ \bot, $ 
   \item
      $ t_{S4}( \bot, 0 ) $ equals $ \top, $ 
   \item
      for a propositional symbol $ \avarprop, $ \
      $ t_{S4}(\avarprop, 1 ) $ equals $ \Box \avarprop, $ 
   \item
      for a propositional symbol $ \avarprop, $ 
      \ $ t_{S4}(\avarprop, 0 ) $ equals
      $ \Diamond \neg \avarprop, $ 
   \item
      $ t_{S4}( \aformulabis \wedge \aformulabis', 1 ) $ equals
      $ t_{S4}( \aformulabis, 1) \wedge 
        t_{S4}( \aformulabis', 1), $ 
   \item
      $ t_{S4}( \aformulabis \wedge \aformulabis', 0) $ equals
      $ t_{S4}( \aformulabis, 0) \vee 
        t_{S4}( \aformulabis', 0), $ 
   \item
      $ t_{S4}( \aformulabis \vee \aformulabis', 1) $ equals
      $ t_{S4}( \aformulabis, 1 ) \vee 
        t_{S4}( \aformulabis', 1 ), $ 
   
   \item
      $ t_{S4}( \aformulabis \vee \aformulabis', 0) $ equals
      $ t_{S4}( \aformulabis, 0 ) \wedge
        t_{S4}( \aformulabis', 0 ), $ 

   \item
      $ t_{S4}( \aformulabis \rightarrow \aformulabis', 1 ) $ equals
      $ \Box ( \ t_{S4}( \aformulabis, 0 ) \vee
                 t_{S4}( \aformulabis', 1 ) \ ), $ 
   \item
      $ t_{S4}( \aformulabis \rightarrow \aformulabis', 0 ) $ equals
      $ \Diamond ( \ t_{S4}( \aformulabis, 1 ) \wedge 
                     t_{S4}( \aformulabis', 0 ) \ ). $ 
\end{itemize}
\end{definition}

It is easily checked that $ t_{S4}(\aformula,\pi) \Leftrightarrow
 \neg  t_{S4}(\aformula, 1 - \pi)$ is a theorem of modal logic K (and of
 modal logic S4 {\em a fortiori}). 
Using this, it is easily checked that $ t_{S4}(\aformula,0) $ 
equals the negation normal form of $ \neg t_{S4}(\aformula). $

Logic S4 has the frame condition that the accessibility relation
should be reflexive and transitive. This can be expressed 
by a semi-Thue system $ \asystem $ by 
taking a singleton alphabet $ \aalphabet = \set{ \aletter }, $ 
and $ \alang_{\asystem}(a) = \aletter^{*}. $ 
This language can be easily recognized by a one-state regular automaton
$ \aautomaton = ( \set{q}, q, \set{q}, \set{ (q,a,q) } ). $
A translation $ t_{\aautomaton}(\alpha, \aformulater) $ 
(Definition~\ref{definition-translation-automaton}) will introduce one unary predicate 
$ {\bf q}_{\aformulater}. $ 
We are now ready to define the translation. 
It introduces
the following symbols: 
\begin{itemize}
\itemsep 0 cm
\item 
   One binary relation $ \relpred{} $ (interpreted as the S4 reflexive 
   and transitive relation);
\item 
   A unary predicate symbol $\proppred{p}$, for every propositional variable 
   $ \avarprop $ in $\aformula$. This symbol serves two purposes at the
   same time: It is the unary predicate symbol representing the
   unique state of $ \aautomaton $ in 
   $ t_{\aautomaton}(\alpha, t( \avarprop, \alpha, \beta,0 )), $ and 
   also the translation of $ \avarprop $ itself. 
\item 
   A unary predicate symbol 
   $ \statepred{\aformulabis \rightarrow \aformulabis'} $, for every 
   subformula of $ \aformula $ that occurs negatively 
   and that has form $\aformulabis \rightarrow \aformulabis'. $ 
   This is the unary predicate needed for representing the state
   of $ \aautomaton $ in the translation 
   $ t_{\aautomaton}( \alpha, 
           t( \ \Box( \ t_{S4}(\aformulabis,0) \vee  
                        t_{S4}(\aformulabis', 1) \ )). $ 
\end{itemize}

\begin{definition} 
   The translation function 
   $t(\aformula, \alpha, \beta, \pi )$ is defined by recursion on
   the subformulae of $\aformula$, for $ \pi \in \{ 0, 1 \}. $  
   \begin{itemize}
   \item $ t(\perp, \alpha, \beta,1) $ equals $ \perp $, 
   \item $ t(\perp, \alpha, \beta,0) $ equals $ \top$,
   \item $ t(\aformulabis \wedge \aformulabis', \alpha, \beta,1) $ equals 
         $ t(\aformulabis, \alpha, \beta,1) \wedge   
           t(\aformulabis', \alpha, \beta,1) $,
   \item $ t(\aformulabis \wedge \aformulabis', \alpha, \beta,0) $ equals 
         $  t(\aformulabis, \alpha, \beta,0) \vee 
            t(\aformulabis', \alpha, \beta,0) $,
   \item $ t(\aformulabis \vee \aformulabis', \alpha, \beta,1) $ equals 
         $ t(\aformulabis, \alpha, \beta,1) \vee 
           t(\aformulabis', \alpha, \beta,1) $,
   \item $ t(\aformulabis \vee \aformulabis', \alpha, \beta,0) $ equals 
         $ t(\aformulabis, \alpha, \beta,0) \wedge 
           t(\aformulabis', \alpha, \beta,0) $,
   \item $ t(\aformulabis \rightarrow \aformulabis', \alpha, \beta,1) $ equals 
         the conjunction 
         \[ \statepred{\aformulabis \rightarrow \aformulabis'}(\alpha) 
             \wedge 
          \forall \alpha \beta \ [ \ \relpred{}(\alpha,\beta) \rightarrow 
                  \statepred{\aformulabis \rightarrow \aformulabis'}(\alpha) 
                                        \rightarrow
                  \statepred{\aformulabis \rightarrow \aformulabis'}(\beta) \ ] 
             \ \wedge \]
         \[ \forall \alpha \ [ \ 
               \statepred{\aformulabis \rightarrow \aformulabis'}(\alpha) 
                            \rightarrow
               t(\aformulabis,  \alpha, \beta, 0 ) \vee
               t(\aformulabis', \alpha, \beta, 1 ) \ ]. \]
   \item $t(\aformulabis \rightarrow \aformulabis', \alpha, \beta,0) $ equals 
          $ \exists \beta \ [ \ \relpred{}(\alpha,\beta)  \wedge 
          t(\aformulabis, \beta, \alpha, 1) 
          \wedge 
          t(\aformulabis', \beta, \alpha, 0) \ ], $
   \item $t(\avarprop, \alpha, \beta,1) $ equals 
         \[ \proppred{\avarprop}(\alpha) \wedge
             \forall \alpha \beta \ [ \ \relpred{}(\alpha,\beta) \rightarrow
                     \proppred{\avarprop}(\alpha) \rightarrow 
                     \proppred{\avarprop}(\beta) \ ],
          \] 
   \item $ t(\avarprop, \alpha, \beta,0) $ equals 
         $ \exists \beta \ [ \ \relpred{}(\alpha,\beta)  \wedge
          \neg \proppred{\avarprop}(\beta) \ ]. $
   \end{itemize}
\end{definition}

We write $T(\aformula)$ to denote 
$ t(\aformula, \alpha, \beta, 0). $ 
Using Theorem \ref{theorem-main-translation} for S4 and 
G\"odel's translation from IPL into S4, one can easily show the following:

\begin{theorem} 
   $\aformula$ is intuitionistically valid iff 
   $T(\aformula)$ is $\GF^2$ unsatisfiable. 
\end{theorem}

\begin{proof}
Indeed, one can easily show that $T(\aformula)$ is satisfiable iff
$T_{\asystem}(t_{S4}(\aformula,0))$ is satisfiable
where $T_{\asystem}$ is defined for the semi-Thue system for S4. 
By Theorem~\ref{theorem-main-translation}, $t_{S4}(\aformula,0)$ is
S4-unsatisfiable
iff  $T(\aformula)$ is $\GF^2$ unsatisfiable. However,
$t_{S4}(\aformula,0)$ is equivalent to $\neg t_{S4}(\aformula,1)$
and $t_{S4}(\aformula,1)$ is S4 valid iff 
$\aformula$ is intuitionistically valid. 
Hence, $\aformula$ is intuitionistically valid iff 
   $T(\aformula)$ is $\GF^2$ unsatisfiable. 
\end{proof}

The translation from IPL into $\GF^2$ can be extended in a similar way 
to various intuitionistic modal logics  from~\cite{Wolter&Zakharyaschev97}.
This allows us to get a uniform decidability proof for such logics even though
in the case of plain IPL the translation is not the tightest one since
IPL provability is \pspace-complete \ whereas $\GF^2$ is \exptime-complete.

\section{Concluding Remarks}
\label{section-conclusion}

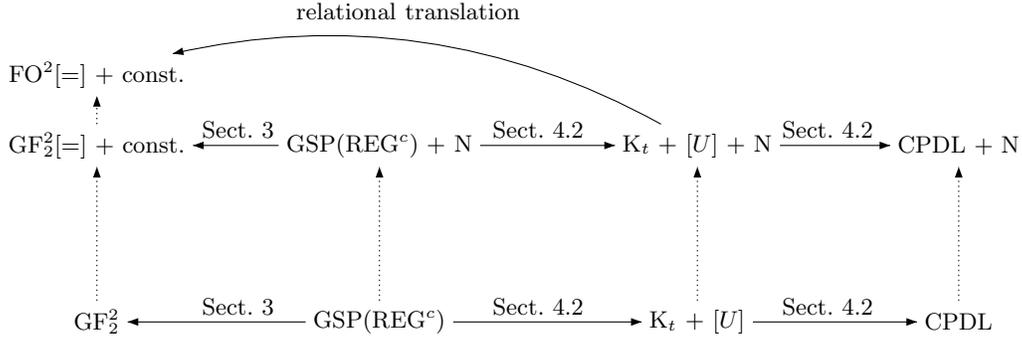
\begin{figure}[htbp]
\centering
{\setlength{\unitlength}{0.94mm}
\begin{picture}(132,45)(2,0)

\gasset{Nframe=n,Nadjust=w,Nh=6,Nmr=0}

\node(GF22)(05,05){$\GF^2_2$}
\node(GF22C)(05,30){$\GF^2_2[=]$ + const.}
\node(FO2C)(05,40){$\FO^2[=]$ + const.}

\node(GSPREG)(45,05){GSP(REG$^c$)}
\node(GSPREGN)(45,30){GSP(REG$^c$) + N}

\node(KU)(90,05){K$_t$ + $\bo{U}$}
\node(KUN)(90,30){K$_t$ + $\bo{U}$ + N}

\node(CPDL)(127,05){CPDL}
\node(CPDLN)(127,30){CPDL + N}

{\gasset{dash={0.2 0.5}0}
\drawedge(GF22,GF22C){}
\drawedge(GF22C,FO2C){}
\drawedge(GSPREG,GSPREGN){}
\drawedge(KU,KUN){}
\drawedge(CPDL,CPDLN){}
}

\drawedge[ELside=r](GSPREG,GF22){Sect.~\ref{section-2gf-translation}}
\drawedge[ELside=r](GSPREGN,GF22C){Sect.~\ref{section-2gf-translation}}

\drawedge(GSPREG,KU){Sect.~\ref{section-ku}}
\drawedge(GSPREGN,KUN){Sect.~\ref{section-ku}}


\drawedge[curvedepth=-10,ELside=r](KUN,FO2C){relational translation}
\drawedge(KUN,CPDLN){Sect.~\ref{section-ku}}
\drawedge(KU,CPDL){Sect.~\ref{section-ku}}

\end{picture}}
\caption{Logarithmic space transformations}
\label{figure-translations}
\end{figure}

Fig.~\ref{figure-translations} contains logarithmic space transformations
 from GSP(REG$^c$) into  logics  such as
$\GF^2_2$ ($\GF^2$ restricted to predicate symbols of arity at most 2),
CPDL, and
K$_t$ + $\bo{U}$ (possibly augmented with nominals, extensions noted with ``+ N'').
In a sense, although  the main contribution of the paper consists in
designing a simple logarithmic space transformation from GSP(REG$^c$)
into $\GF^2_2$ by simulating the behaviour of finite automata, 
Fig.~\ref{figure-translations} shows other encodings of 
GSP(REG$^c$) which could be effectively used to mechanize modal logics
captured by GSP(REG$^c$) (see examples of such logics throughout the paper).

The encoding we used  is 
reminiscent to the propagation of formula in tableaux calculi
(see 
e.g.~\cite{Gore99,Massacci00,Castilho&Farinas&Gasquet&Herzig97,Farinas&Gasquet02}) 
and the study of such a relationship may be worth being pursued.
The study of the computational behaviour of the translation to mechanize modal
logics using for instance~\cite{deNivelle&PrattHartmann01} is also an
interesting direction for future work.

Additionally, our work allows us
to  answer positively to some questions left
open in \cite{Demri01}.
Typically, we provide evidence that the first-order fragment to translate into
the regular grammar logics with converse is simply $\GF^2$: no need for 
first-order fragment augmented with fixed-point operators.
Moreover, we characterize the complexity of such logics
and we illustrate how the map can be extended for other non-classical logics
 including nominal tense logics and intuitionistic
   logic.  \\

\noindent
We list a few open problems that we believe are worth investigating.
\begin{enumerate}
\itemsep 0 cm
\item Although regular grammar logics (with converse)
can be viewed as fragments of propositional dynamic logic, it
remains open whether the full PDL can be translated into $\GF^2$ with a 
similar, almost-structure preserving transformation.
We know that there exists a logarithmic space transformation, but we do not 
want to use first principles on Turing machines.
\item How to design a \pspace \
fragment of $\GF^2$ in which the following modal logics can be
naturally embedded: S4, S4$_t$ (S4 with past-time operators), Grz, and G? 
(to quote a few modal logics in \pspace, 
see e.g.~\cite{Chagrov&Zakharyaschev97}).
\item Can our translation method be extended to first-order modal logics?
\item Can it be extended to first-order intuitionistic logic?
\item Finally, a further comparison of the recent
  work~\cite{Hustadt&Schmidt03} with ours should be carried out. 
\end{enumerate}

{\small

}

\end{document}